\documentclass[prl,longbibliography,twocolumn,preprintnumbers,notitlepage,amsmath,amssymb,superscriptaddress]{revtex4-2}

\usepackage{amsmath}
\usepackage{amssymb}
\usepackage{amsbsy}
\usepackage{graphicx}
\usepackage{color}

\usepackage[normalem]{ulem}

\usepackage{mathrsfs}

\usepackage{enumerate}
\usepackage{mathtools}

\newcommand{\moy}[1]{\left\langle #1 \right\rangle}
\newcommand{\dd}[0]{\mathrm{d}}

\def\rt{\tilde{\rho}}

\def\qt{\tilde{q}}
\def\pt{\tilde{p}}
\def\Dt{\tilde{D}}
\def\st{\tilde{\sigma}}

\newcommand{\dt}[2]{\ensuremath{\frac{\dd #1}{\dd #2}}}

\def\e{\mathrm{e}}

\DeclareMathOperator{\erfc}{erfc}

\definecolor{darkblue}{rgb}{0,0,0.6}
\definecolor{darkred}{rgb}{0.6,0,0}
\usepackage[colorlinks=true,urlcolor=darkblue,citecolor=darkblue,linkcolor=darkred]{hyperref}

\begin{document}

\title{Driven Tracer in the Symmetric Exclusion Process: Linear Response and Beyond}

\author{Aur\'elien Grabsch}
\thanks{These two authors contributed equally.}
\affiliation{Sorbonne Universit\'e, CNRS, Laboratoire de Physique Th\'eorique de la Mati\`ere Condens\'ee (LPTMC), 4 Place Jussieu, 75005 Paris, France}

\author{Pierre Rizkallah}
\thanks{These two authors contributed equally.}
\affiliation{Sorbonne Universit\'e, CNRS, Physico-Chimie des \'Electrolytes et Nanosyst\`emes Interfaciaux (PHENIX), 4 Place Jussieu, 75005 Paris, France}

\author{Pierre Illien}
\affiliation{Sorbonne Universit\'e, CNRS, Physico-Chimie des \'Electrolytes et Nanosyst\`emes Interfaciaux (PHENIX), 4 Place Jussieu, 75005 Paris, France}

\author{Olivier B\'enichou}
\affiliation{Sorbonne Universit\'e, CNRS, Laboratoire de Physique Th\'eorique de la Mati\`ere Condens\'ee (LPTMC), 4 Place Jussieu, 75005 Paris, France}

\begin{abstract}

Tracer dynamics in the Symmetric Exclusion Process, where hardcore particles diffuse on an infinite one-dimensional lattice, is a paradigmatic model of anomalous diffusion. While the equilibrium situation has received a lot of attention, the case where the tracer is driven by an external force, which provides a minimal model of nonequilibrium transport in confined crowded environments, remains largely unexplored. Indeed, the only available analytical results concern the means of both the position of the tracer and the lattice occupation numbers in its frame of reference, and  higher-order moments but only in the high-density limit. 
Here, we provide a general hydrodynamic framework that allows us to determine the first cumulants of the bath-tracer correlations and of the tracer's position
in function of the driving force, up to quadratic order (beyond linear response).
This result constitutes the first determination of the bias-dependence of the variance of a driven tracer in the SEP for an arbitrary density. The framework presented here can be applied, beyond the SEP, to more general configurations of a driven tracer in interaction with obstacles in one dimension.

\end{abstract}

\maketitle

\emph{Introduction.---} Single-file transport, corresponding to the diffusion of particles in narrow channels, so that they cannot bypass each other, is observed in various physical, chemical or biological systems, such as zeolites, colloidal suspensions, or carbon nanotubes~\cite{Hahn:1996,Wei:2000,Lin:2005,Cambre:2010}. In this confined geometry, a tracer displays an anomalous subdiffusive behaviour, which has been observed by passive microrheology~\cite{Hahn:1996,Wei:2000,Lin:2005}.
The Symmetric Exclusion Process (SEP) is a paradigmatic model of such single-file diffusion \cite{Chou2011a,Mallick2015}, which has been the object of several recent and important developments \cite{Imamura2017,Poncet2021,Grabsch2021,Mallick2022a}. In this model, particles perform symmetric random walks in continuous time on an infinite one-dimensional lattice, with the constraint that there can only be one particle per site. Characterising the anomalous dynamics of a tracer in this many-body problem has been the subject of a number of theoretical works \cite{Arratia1983,Illien2013a,Hegde2014, Krapivsky2014,Krapivsky2015,Imamura2017,Imamura2021,Poncet2021, Grabsch2021}.
These results are part of a context of intense activity around exact solutions for one-dimensional interacting particle systems  \cite{Krajenbrink2021,Bettelheim2022,Mallick2022a,Derrida2009a,Derrida2009}.

\begin{figure}
\begin{center}
    \includegraphics[width=0.8\columnwidth]{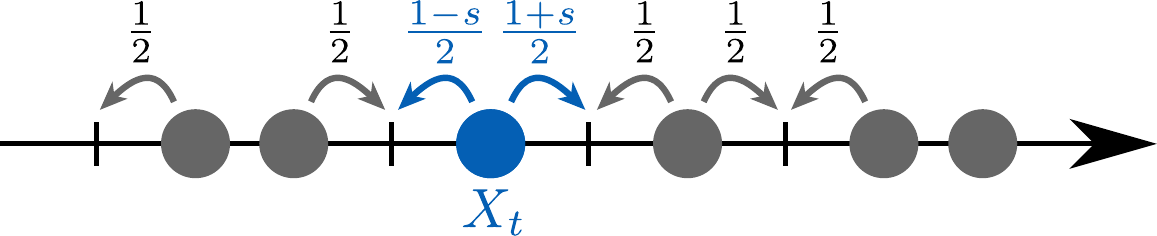}
\end{center}
\caption{The Symmetric Exclusion Process (SEP) with a driven tracer (blue) at position $X_t$ (see section \textit{Model}).}
\label{fig:SEPbias}
\end{figure}

An important extension of tracer diffusion in the SEP concerns the case where the tracer is submitted to an external driving force \cite{Ferrari1985} (see Fig.~\ref{fig:SEPbias}). This situation is encountered for instance in 
active microrheology, which is a technique used to probe the properties of living or colloidal systems by forcing the displacement of a tracer through the medium~\cite{Habdas:2004,Bausch:1998}. More generally, it constitutes a minimal one-dimensional model for nonequilibrium transport in confined crowded environments, which has received a growing attention \cite{Leibovich2013, Lizana2010} (see also \cite{Benichou2018,Leitmann2017,Leitmann2013,Cividini2016a,Cividini2016} for related models combining tracer driving and bath-induced crowding).
This model allows to go beyond the usual Gaussian approximation and characterize the non Gaussian fluctuations, as well as the nonlinear effects of the driving force on the tracer.
The only analytical results at arbitrary density concern the means of both the position of the tracer and the lattice occupation numbers in its frame of reference (i.e. the density profiles) \cite{Burlatsky1996,Burlatsky1992,Landim1998}, which have recently been determined also on finite periodic systems \cite{Lobaskin2020,Ayyer}. Since the seminal works \cite{Burlatsky1996,Burlatsky1992,Landim1998} that date back to almost three decades, the results concerning higher-order cumulants have been limited to the high-density limit  \cite{Illien2013a, Poncet2021a}, and to specific situations \cite{Landim2000}\footnote{The specific situation in~\cite{Landim2000} corresponds to the case where the driving force imposed on the tracer is compensated by a step of density resulting in a vanishing mean position}. At arbitrary density, even the determination of the variance of the position of the tracer, which is crucial to quantify its fluctuations, remains a fully open problem.

In this Letter, we fill this gap and provide a general hydrodynamic framework that allows us to determine at long time bath-tracer density profiles and cumulants of the tracer position at linear order in the driving force and at arbitrary density.  We also go beyond linear response by determining the second cumulant of the tracer position and the corresponding density profile at second order in the driving force. We thus provide the first non-trivial contribution of the driving force to the variance of the tracer position at arbitrary density.

\emph{Model.---} Each site of an infinite 1d lattice is initially occupied by a particle with probability $\rho$. Particles perform symmetric continuous-time random walks  with half unit jump rate onto each nearest neighbor, and with the hard-core constraint that there is at most one particle per site. A tracer,  of position $X_t$  at time $t$, is initially at the origin, and is the only particle to experience a driving force, which results in asymmetric jump rates, namely $(1+s)/2$ to the right and $(1-s)/2$ to the left. The parameter $s$ quantifies the asymmetry and will be called the bias. The bath particles are described by the set of occupation numbers  $\eta_r(t)$ of each site $r\in\mathbb{Z}$ of the lattice at time $t$, with $\eta_r(t) = 1$ if the site is occupied and $\eta_r(t) = 0$ otherwise.

We first derive the hydrodynamic limit of the problem, by extending to the case of a driven tracer the approach we developed to study a symmetric tracer in \cite{Poncet2021,Grabsch2021}. We consider the cumulant generating function (CGF) of the position of the tracer: $    \psi(\lambda,t) = \ln \moy{\e^{\lambda X_t}} = \sum_{n=0}^\infty \frac{\lambda^n}{n!} \kappa_n$,  where the $\kappa_n$ are the cumulants of the position of the tracer. Its time evolution is deduced from the master equation given in the Supplementary Material \cite{SM}, and reads
\begin{equation} \label{eq:link_psi_w}
\frac{\mathrm{d}\psi}{\mathrm{d} t} = \frac{1}{2}\sum_{\nu=\pm 1} \left[ (1+\nu s)(\mathrm{e}^{\nu \lambda} - 1)(1-w_\nu) \right],
\end{equation}
where we have denoted $w_r(t) = {\moy{\eta_{X_t+r}\e^{\lambda X_t}}}/{\moy{\e^{\lambda X_t}}}$. We call $w_r$ the generalised density profile generating function, since by expanding it in powers of $\lambda$ it generates all correlation functions between the displacement of the tracer and the density of bath particles at a distance $r$ from the tracer (represented by the occupation number $\eta_{X_t+r}$): $ w_r(t) = \sum_{n \geq 0} \frac{\lambda^n}{n!}
    \moy{\eta_{X_t+r} X_t^n}_c$
with $\moy{\cdots}_c$ the joint cumulants. For instance, at order $1$ in $\lambda$, $\moy{\eta_{X_t+r} X_t}_c = \moy{\eta_{X_t+r} X_t} - \moy{\eta_{X_t+r}} \moy{ X_t}$.
Beyond controlling the displacement of the tracer [Eq.~\eqref{eq:link_psi_w}] and measuring the response of the bath of particles, these profiles $w_r$ are key quantities in the SEP since, in the symmetric case $s=0$, they satisfy a strikingly simple closed equation \cite{Grabsch2021}.

In the hydrodynamic limit of large time and large distances, the different observables have the scalings,
\begin{equation}
    \label{eq:hydro_scalings}
    \psi(\lambda,t) \underset{t \to \infty}{\simeq}
    \hat{\psi}(\lambda) \: \sqrt{2t}
    \:,
    \quad
    w_r(t) \underset{t \to \infty}{\simeq}
    \Phi \left(
    v = \frac{r}{\sqrt{2t}}
    \right)
\end{equation}
where we have omitted the dependency in $\lambda$ of $\Phi$ for simplicity. These scalings have been shown to hold in the symmetric case~\cite{Imamura2017,Grabsch2021,Mallick2022a}, and in the biased case~\cite{Landim1998} at lowest orders in $\lambda$ for arbitrary density and at all orders in the high density limit. Here, based on numerical observations, we extend  Eq.~\eqref{eq:hydro_scalings} to all orders in $\lambda$. From Eq.~\eqref{eq:link_psi_w}, these scalings imply the boundary condition
\begin{equation}
    \label{eq:BoundPhi0}
    \sum_{\nu = \pm 1}(1+\nu s) (\e^{\nu \lambda}-1) (1-\Phi(0^\nu)) 
    = 0 \:.
\end{equation}
Another key boundary condition is obtained from the time evolution of $w_{\pm 1}$ deduced from the master equation~\cite{SM},
\begin{equation}
    \Phi'(0^\pm) \pm \frac{2 \hat{\psi}}{\e^{\pm \lambda}-1}
    \Phi(0^\pm)
    \label{eq:BoundPhi0_bis}
    = 0 \:.
\end{equation}
Remarkably,  Eq.\eqref{eq:BoundPhi0_bis} is closed and does not involve higher-order correlation functions.

In contrast, the bulk equation satisfied by $\Phi(v)$ is not closed. 
Thus, to compute this profile, we design another approach \footnote{Note that we became aware of an approach similar to the one presented here while we were finalizing this manuscript \cite{Dandekar2022}. Although both works start from the same hydrodynamic equations, the explicit results by Dandekar and Mallick focus on the high-density limit of the problem, while our results are valid at arbitrary density and were out of reach from available microscopic approaches.} based on a fluctuating hydrodynamic description.

\emph{Macroscopic fluctuation theory (MFT) for a driven tracer.---} This approach relies on MFT, which is a powerful tool to treat the stochastic dynamics of diffusive systems at large scale \cite{Bertini2015}, and to determine the statistics of observables in single-file systems such as the current \cite{Derrida2009a, Mallick2022a} or the position of a symmetric tracer \cite{Krapivsky2014, Krapivsky2015}. 
The MFT expresses the probability of observing a fluctuation of the macroscopic profile $q(x,t)$, representing the density of particles, in terms of a diffusion coefficient $D(\rho)$ and a mobility $\sigma(\rho)$ characterising the system at large scales \cite{Spohn:1991}. Below, we mainly focus on the SEP for which $D(\rho) = 1/2$ and $\sigma(\rho) = \rho(1-\rho)$, but the methodology is general. The case of a driven tracer introduces technical difficulties: (i) the driving force experienced by the tracer creates a discontinuity in the MFT fields at the location of the tracer; (ii) the location of this discontinuity is moving with time.

We circumvent these difficulties by mapping the original problem onto a dual problem where the position of the tracer $X_t$ is translated into a flux at the origin $Q_t$, therefore transforming the moving boundary condition into a static one located at zero~\cite{Evans2005,Rizkallah2022}. A similar approach was used in~\cite{Kundu2016} for a different model.
The dual system is described by new MFT fields $\pt$ and $\qt$,  where $\qt(k,t)$ represents the distance between the particles labelled by the index $k$, which becomes a continuous variable at the hydrodynamic level considered here. These fields obey the following MFT equations (see SM \cite{SM} or \cite{Rizkallah2022} for derivation):
  \begin{subequations}
    \label{eq:MFTpq}
    \begin{align}
      \label{eq:MFT_qt}
      \partial_t \qt 
    &= \partial_k(\Dt(\qt) \partial_k \qt)
    - \partial_k (\st(\qt) \partial_k \pt)
                        \:,
      \\
      \label{eq:MFT_pt}
      \partial_t \pt 
    &= -\Dt(\qt) \partial_k^2 \pt
    - \frac{1}{2}\st'(\qt) (\partial_k \pt)^2 \:,
    \end{align}
  \end{subequations}
which involve the transport coefficients of the dual system $\Dt(\rt) = D(1/\rt)/\rt^2$ and $\st(\rt)= \rt \sigma(1/\rt)$. The initial and final conditions are
\begin{equation}
    \label{eq:sm_bound_pt}
    \pt(k,0) = 
    \int_{\rt}^{\qt(k,0)} \frac{2 \Dt(z)}{\st(z)} \dd z 
    - \lambda \Theta(k)
    \:,
    \quad
    \pt(k,1)
    = -\lambda \Theta(k)
    \:,
\end{equation}
where $\rt=1/\rho$, and $\Theta$ is the Heaviside function. Equations~(\ref{eq:MFTpq},\ref{eq:sm_bound_pt}) are the usual MFT equations, completed here by matching conditions at the origin (reminiscent of the position of the tracer in the original system) which implement the bias~\footnote{The methodology presented here is general, but Eqs.~\eqref{eq:matchingDp} and~\eqref{eq:biais_cond_dual} are written for the SEP.}:
\begin{align}
    \label{eq:matching}
 &    \pt(0^+,t) = \pt(0^-,t) \:, \\
    \label{eq:matchingDp}
  &   (1-s) \partial_k \pt(0^+,t)
    = (1+s) \partial_k \pt(0^-,t) \:, \\
    \label{eq:ContJ0}
 &    \left[ -\Dt(\qt) \partial_k \qt + \st(\qt) \partial_k \pt \right]_{0^-}^{0^+} = 0 \:.
\end{align}
The first two equations originate from the optimization of the MFT action, and the third one comes from the continuity of the current at the origin. The last matching condition is a consequence of Eq. \eqref{eq:BoundPhi0} (see \cite{SM} for details):
\begin{equation} 
    \label{eq:biais_cond_dual}
    (1+s)\left( 1 - \frac{1}{\qt(0^+,t)} \right)
    = (1-s) \left( 1 - \frac{1}{\qt(0^-,t)} \right)
    \:.
\end{equation}
Equations \eqref{eq:MFT_qt}--\eqref{eq:biais_cond_dual} fully determine the dual MFT fields. Finally, the generalized density profiles of the original tracer problem are obtained from these solutions by
\begin{equation}
    \label{eq:Phi_from_qt}
    \Phi\left(v = \frac{y(k)}{\sqrt{2}} \right)
    =  \frac{1}{\qt(k,1)}
    \:,
    \quad
    y(k) = \int_0^k \qt(k',1) \dd k'
    \:.
\end{equation}

This completely sets the problem of a driven tracer in the SEP. However, since there is a priori no explicit solution for arbitrary density and arbitrary bias, this remains formal at this stage. We now go further and propose two lines of investigation of these equations: (i) a numerical resolution for arbitrary sets of parameters; (ii) and  a perturbative expansion, which yields explicit results valid at \emph{arbitrary} density for the first coefficients $\Phi_n(v)$ defined by the expansion of the hydrodynamic limit of the generalized density profiles: $\Phi(v)= \sum_{n=0}^\infty \frac{\lambda^n}{n!} \Phi_n(v)$.

\begin{figure}
\begin{center}
\includegraphics[width=0.49\columnwidth]{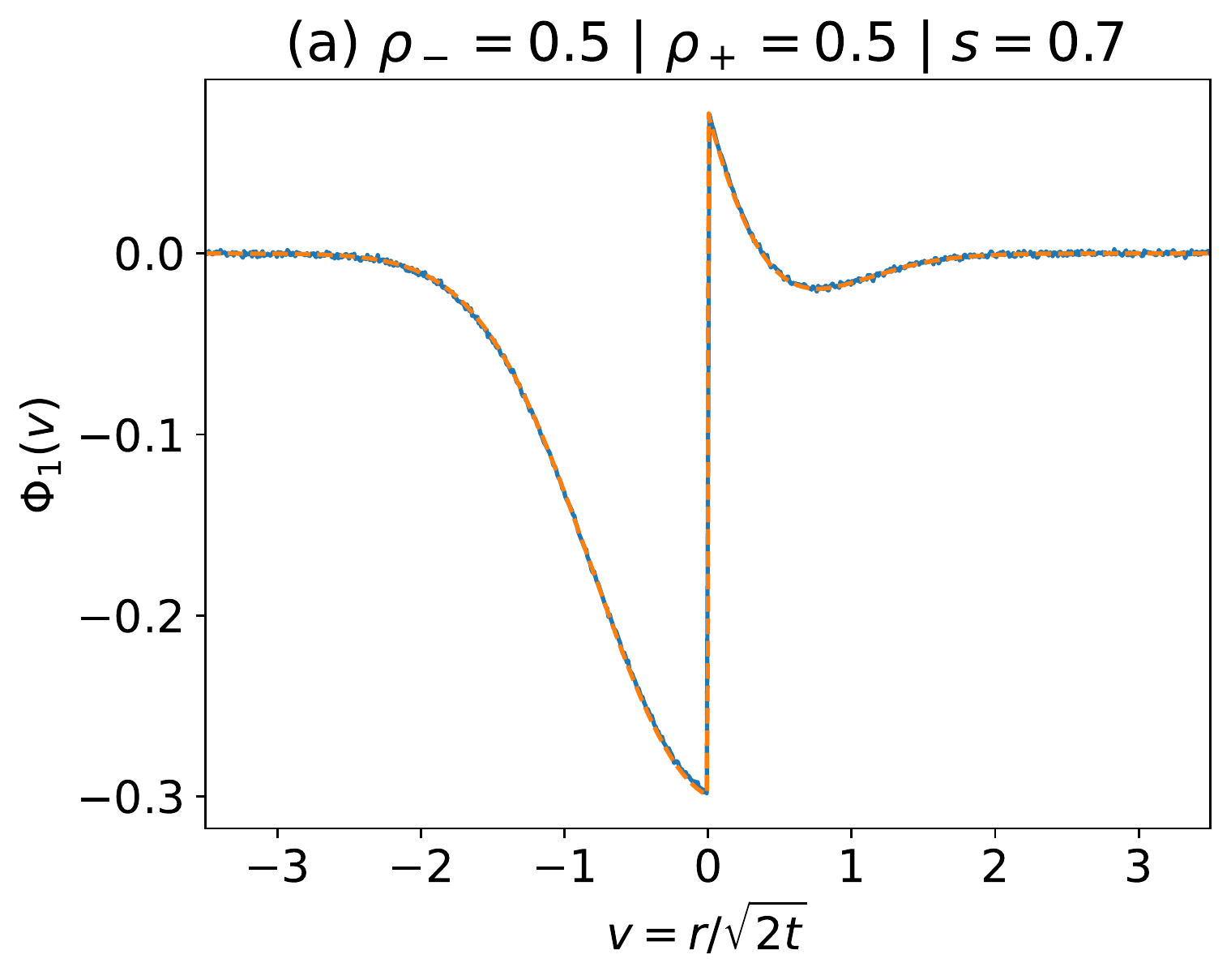} 
\includegraphics[width=0.49\columnwidth]{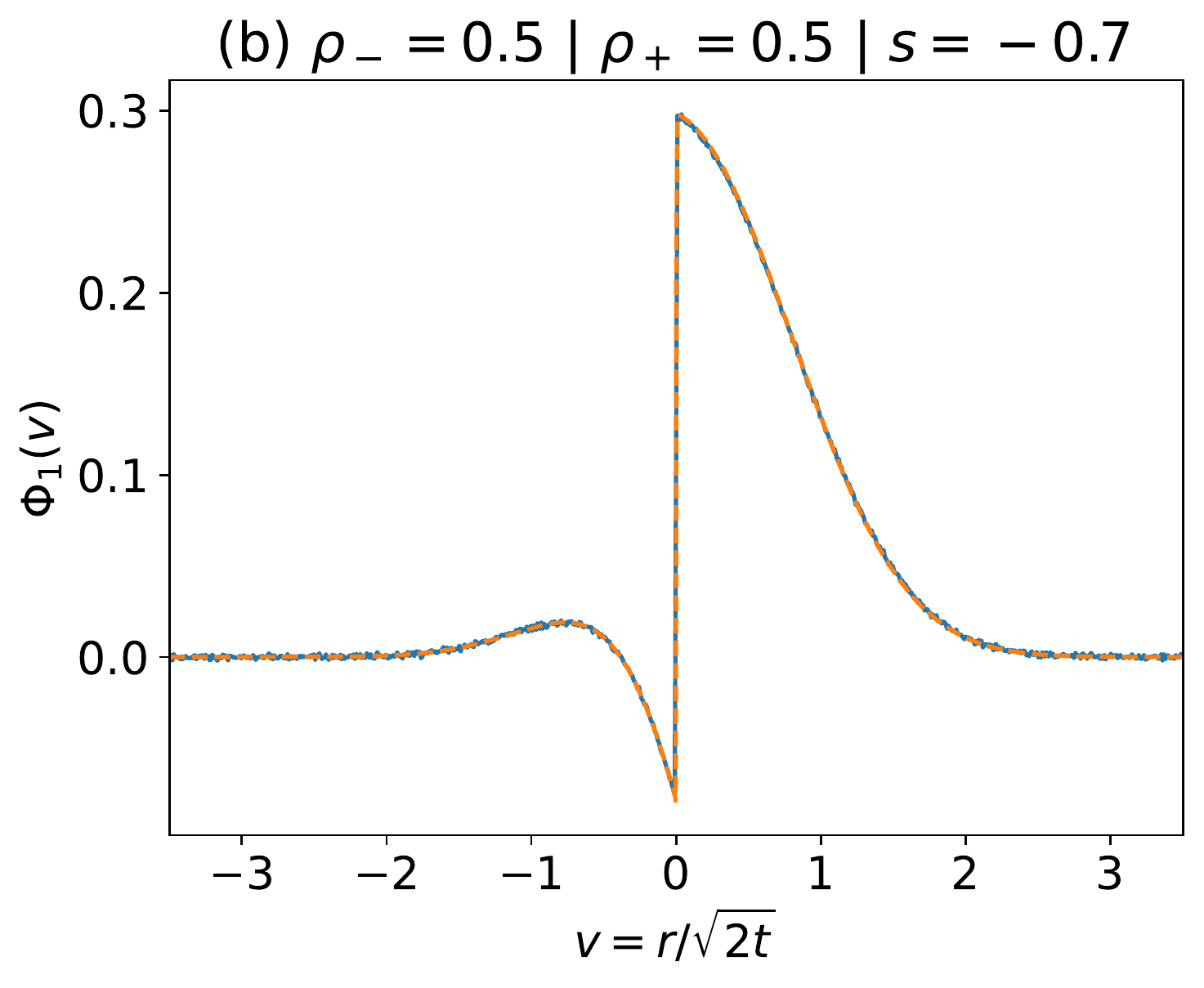} 
\includegraphics[width=0.49\columnwidth]{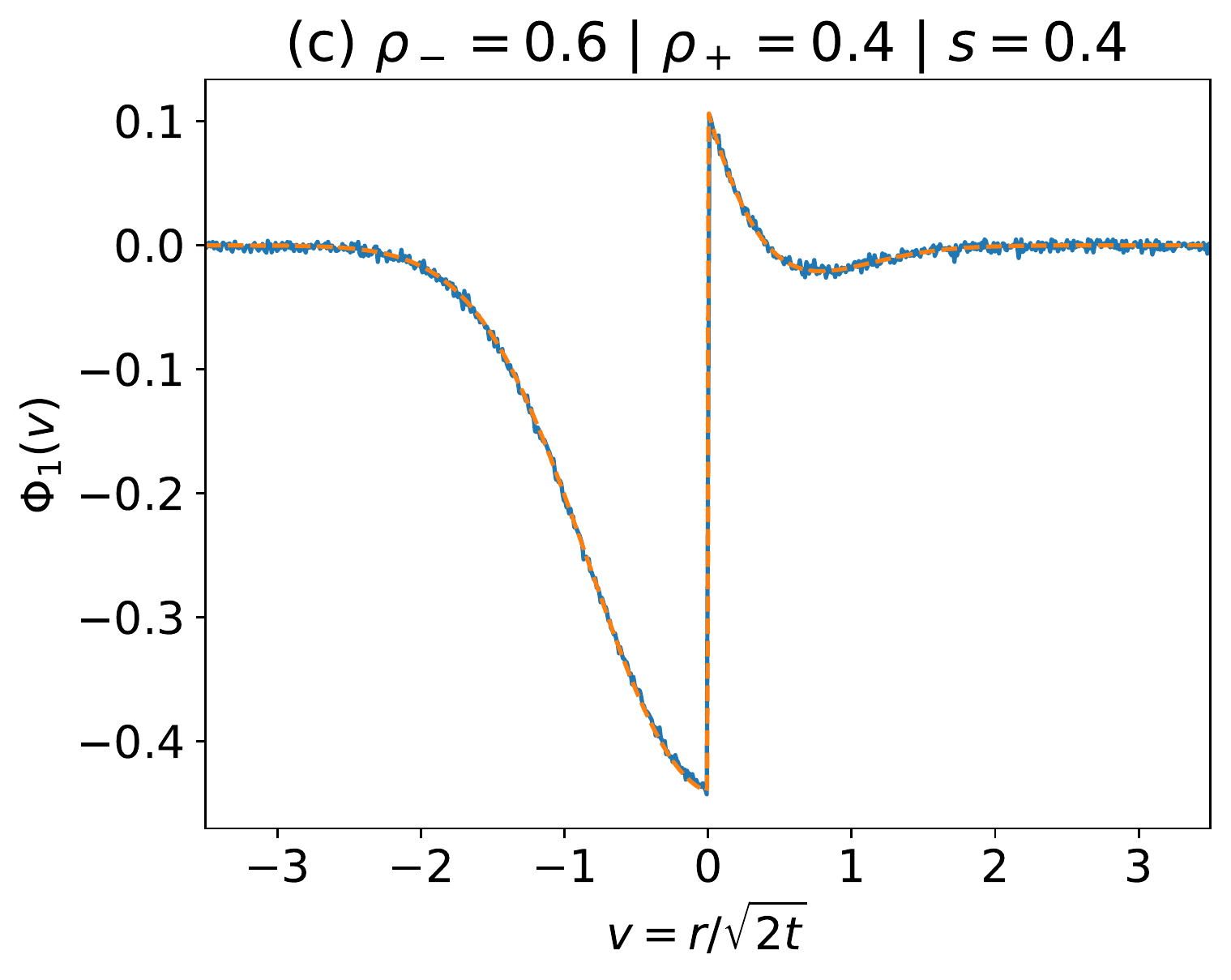} 
\includegraphics[width=0.49\columnwidth]{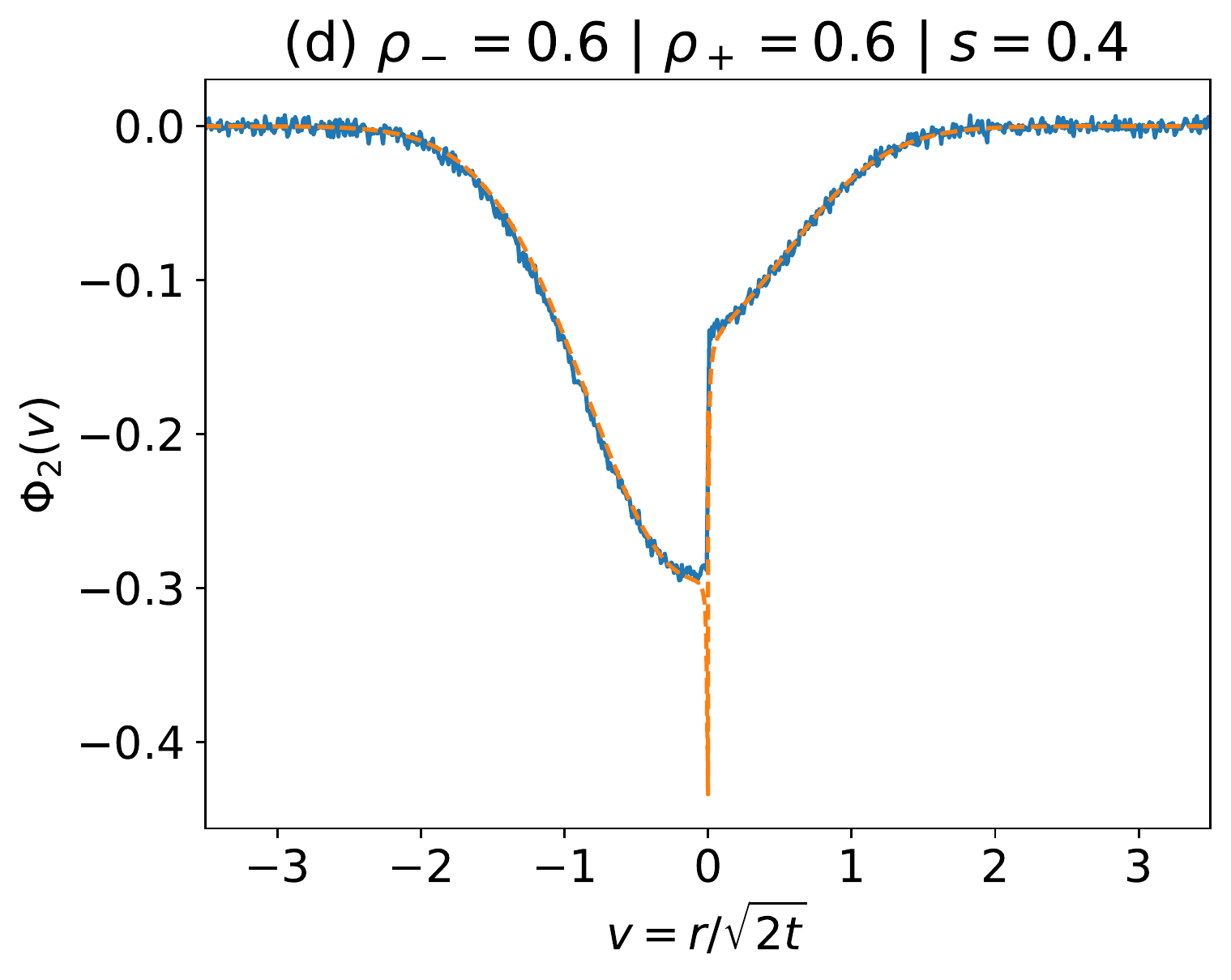} 
\end{center}
\caption{Profiles $\Phi_1$ and $\Phi_2$ obtained by the numerical resolution of the MFT equations \eqref{eq:MFT_qt} and  \eqref{eq:MFT_pt} (orange dashed lines), compared to Monte Carlo simulations (blue solid lines), final time 6000, $10^7$ simulations for (a), (b) and (c) and $9 \cdot 10^7$ for (d), of the SEP with a driven tracer, for various values of the bias and the density. (a) $\Phi_1$ for $\rho=0.5$ and $s=0.7$. (b) $\Phi_1$ for $\rho=0.5$ and $s=-0.7$. (c) $\Phi_1$ for a step density with $\rho_-=0.6$, $\rho_-=0.4$ and $s=0.4$. (d) $\Phi_2$ for $\rho=0.6$ and $s=0.4$. The discrepancy at $v=0$ on panel (d) comes from the numerical errors on $\Phi_1$ near the discontinuity at the origin, which are amplified at the second order $\Phi_2$.}
\label{fig:biasedSEPverif}
\end{figure}

\emph{Numerical resolution.---} We show in Fig.~\ref{fig:biasedSEPverif} the profiles at order $1$ and $2$ in $\lambda$ obtained by the numerical resolution of the MFT equations (see SM \cite{SM} for details), which are in perfect agreement with results from microscopic Monte Carlo simulations (see SM \cite{SM}),  for a broad range of parameters. In particular, we consider strong biases, and densities which are far from the extreme low- and high-density limits.  Note that the approach can be extended to the paradigmatic case where the initial density of particles is step-like ($\rho=\rho_+$ in front of the tracer and $\rho=\rho_-$ behind the tracer) \cite{Imamura2017, Derrida2009}. Finally, the plots show that our MFT procedure  captures non-trivial dependencies of the correlation profiles on the rescaled distance.

\begin{figure}
  \centering 
  \includegraphics[width=0.49\columnwidth]{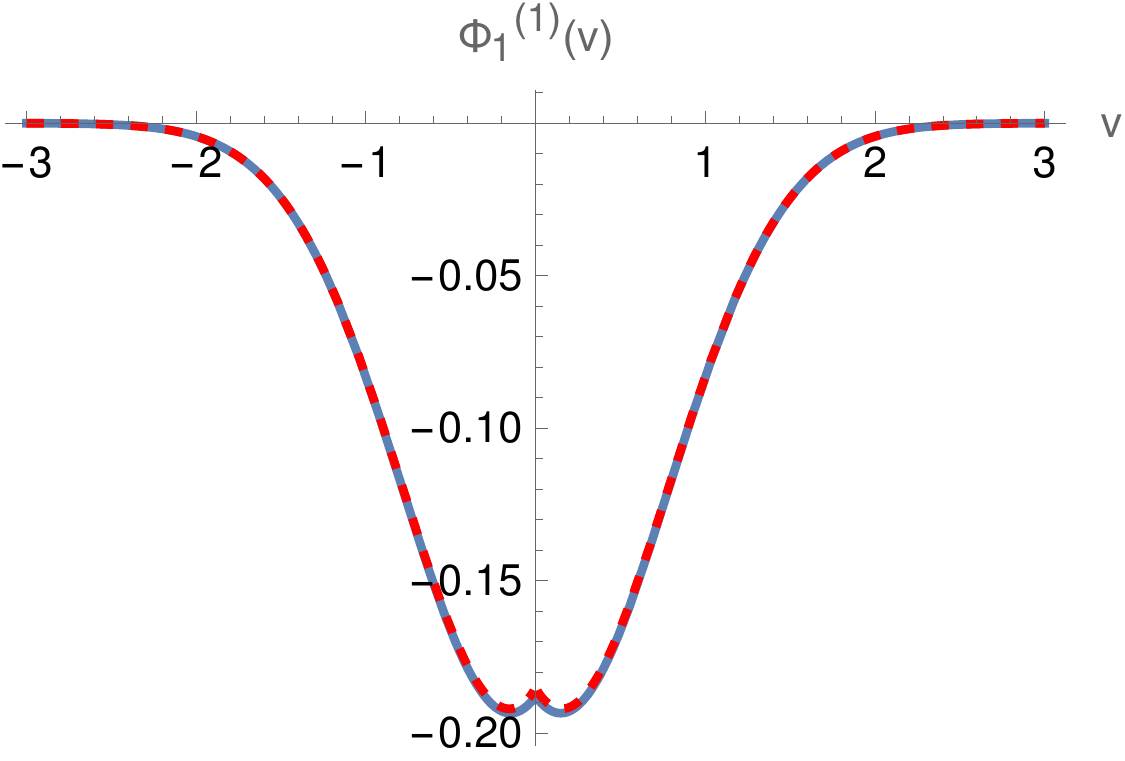}
  \includegraphics[width=0.49\columnwidth]{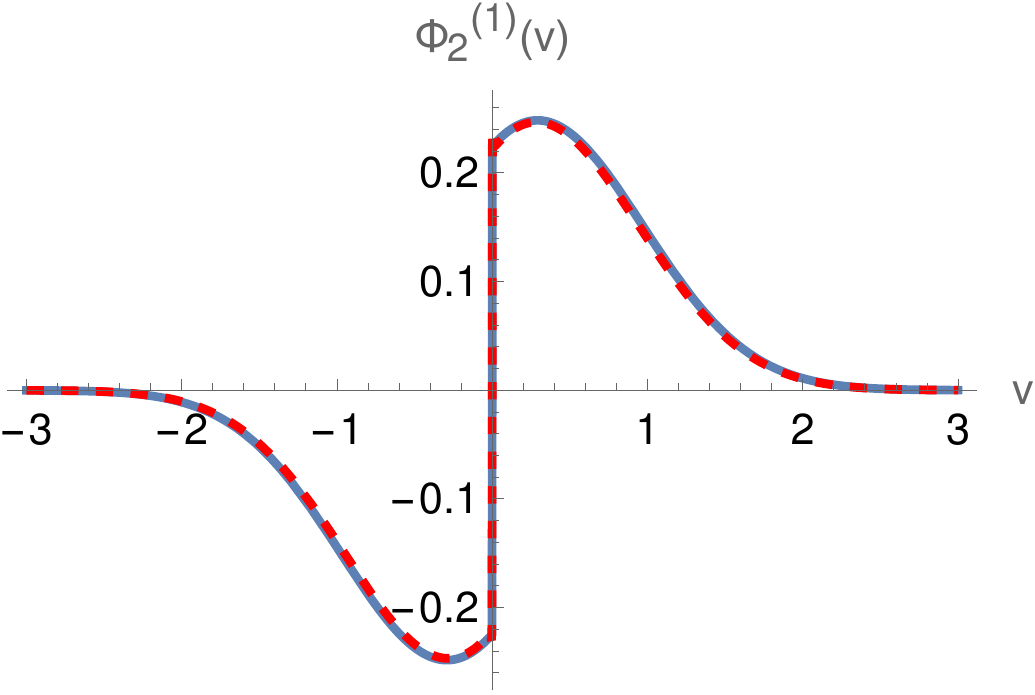}
  \caption{Generalised density profiles $\Phi_n^{(1)}(v)$ at first order in the bias $s$, at density $\rho=0.6$, obtained from the numerical resolution of the MFT equations \eqref{eq:MFT_qt} and  \eqref{eq:MFT_pt} (dashed red lined), compared to the analytical expressions~\eqref{eq:Phi11} and~\eqref{eq:MFTq21biais} (solid blue).  Left: profile $\Phi_1^{(1)}$. Right: profile $\Phi_2^{(1)}$.}
  \label{fig:q21Biais}
\end{figure}

\emph{Linear order in $s$.---} We first note that, for any bias, at zeroth order in $\lambda$, we retrieve the exact results previously obtained for the mean occupation profiles in the frame of reference of the driven tracer \cite{Burlatsky1996, Landim1998}. However, for the next orders ($\Phi_n$ with $n\geq 1$), no explicit analytical solution of the MFT problem at arbitrary density is available. We then resort to an expansion in powers of the bias $s$, and define for each order $n$:
\begin{equation}
    \Phi_n(v) \underset{s\to 0}{=} \Phi_n^{(0)}(v) + s\Phi_n^{(1)}(v) + s^2\Phi_n^{(2)}(v)  +\dots
    \:
\end{equation}
where $\Phi_n^{(0)}$  corresponds to the known symmetric case \cite{Poncet2021,Grabsch2021}. At linear order in the bias $s$, we find \footnote{We give the expressions for $v>0$, as the ones for $v<0$ can be deduced from the symmetry $v \to -v$, $\lambda \to -\lambda$ and $s \to -s$ which imposes $\Phi_n^{(m)}(-v) = (-1)^{n+m}\Phi_n^{(m)}(v)$.}\cite{SM}
\begin{equation}
    \label{eq:Phi11}
  \Phi_1^{(1)}(v)
  = \frac{1-\rho}{2\rho} \left(
    (2-3\rho) \erfc (v) -
    (1-\rho) \frac{6}{\pi} \e^{-v^2}
    \right)
\end{equation}
\begin{widetext}
\begin{multline}
  \label{eq:MFTq21biais}
  \Phi_2^{(1)}(v) =
  \frac{(1-\rho)(1-2 \rho(1-\rho))}{2 \rho^2} \erfc (v)
  + \frac{(1-\rho)^2(4-3\rho)}{\pi \rho^2} \erfc(v)
  - \frac{(1-\rho)^2}{\rho}  \erfc \left( \frac{v}{\sqrt{2}} \right)^2
  \\
  - \frac{(1-\rho)^2}{2 \rho^2} G (\sqrt{2} \: v) 
  + \frac{8(1-\rho)^3}{\pi^{3/2} \rho^2} v \: \e^{-v^2}
  - \frac{4(1-\rho)^2(1-2\rho)}{\pi \rho^2} \e^{-v^2}
  - \frac{(1-\rho)^2}{\pi^{3/2} \rho^2}  v \: \e^{-\frac{v^2}{2}} \: \mathrm{K}_0 \left( \frac{v^2}{2} \right)
  \:,
\end{multline}
\end{widetext}
where
$  G(x) = \frac{1}{\pi} \sqrt{\frac{2}{\pi}}
  \int_x^\infty \e^{-z^2/4} \mathrm{K}_0 \left( \frac{z^2}{4} \right) \dd z$,
and $\text{K}_0$ is a modified Bessel function of zeroth order. A key point is that, contrary to the first order in $\lambda$, $\Phi_2^{(1)}$ is a {\it non-analytic} function of the rescaled distance $v$, displaying a logarithmic singularity at the origin. This appears to be a specificity of the driven case, since, in the symmetric case, all $\Phi_n$ are analytical functions of the rescaled distance \cite{Grabsch2021}.  The functions $ \Phi_1^{(1)}(v)$ and $ \Phi_2^{(1)}(v)$ are plotted in Fig.~\ref{fig:q21Biais} and display perfect agreement with the numerical resolution of the MFT equations. The profile $\Phi_1(v)$ measures the correlation between the density at a rescaled distance $v$ from the tracer, and the position of the tracer~\cite{Poncet2021}. When there is no driving force, $\Phi_1^{(0)}(v>0) > 0$, therefore a fluctuation of $X_t$ towards the right is correlated with an increase of the density in front of the tracer, indicating an accumulation of particles in front of the tracer. Here, we find that the linear correction to these correlations due to the presence of the drive, $\Phi_1^{(1)}(v)$, is negative, indicating that a positive driving force reduces these correlations, while a negative drive increases them.

\begin{figure}
  \centering
  \includegraphics[width=0.49\columnwidth]{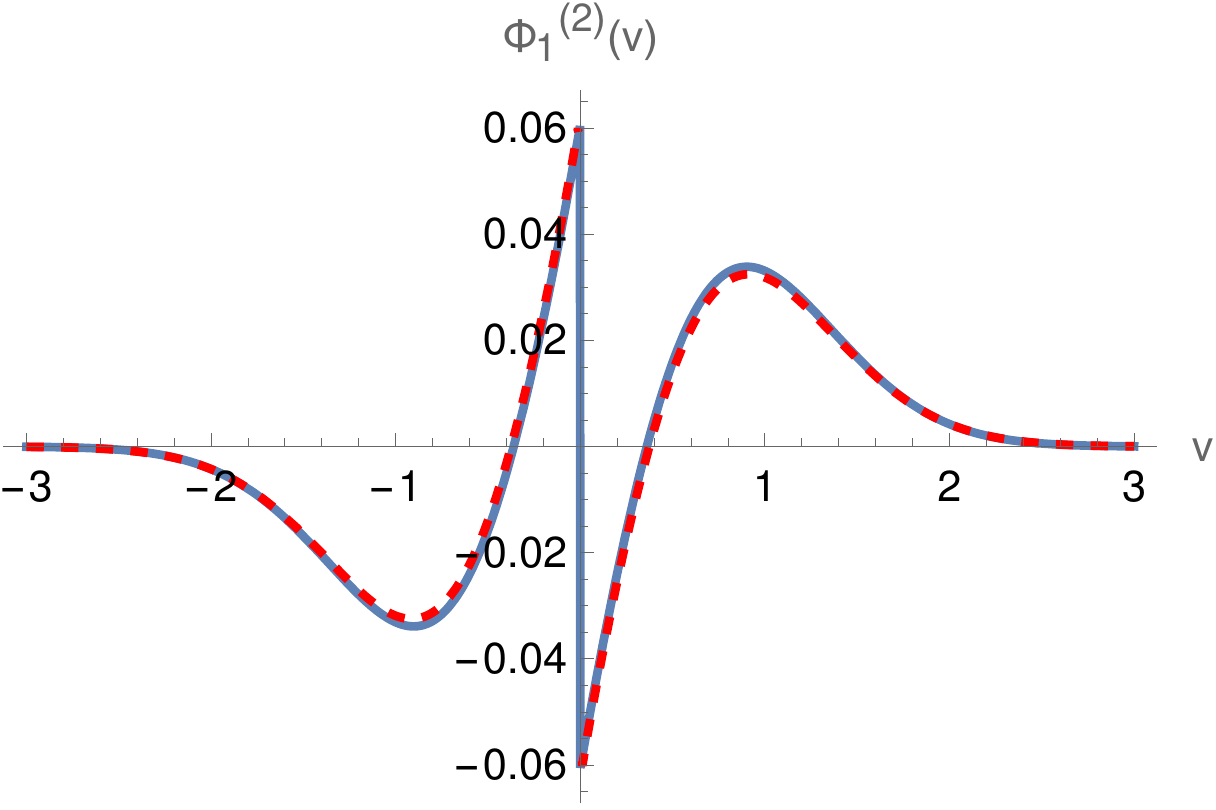}
  \includegraphics[width=0.49\columnwidth]{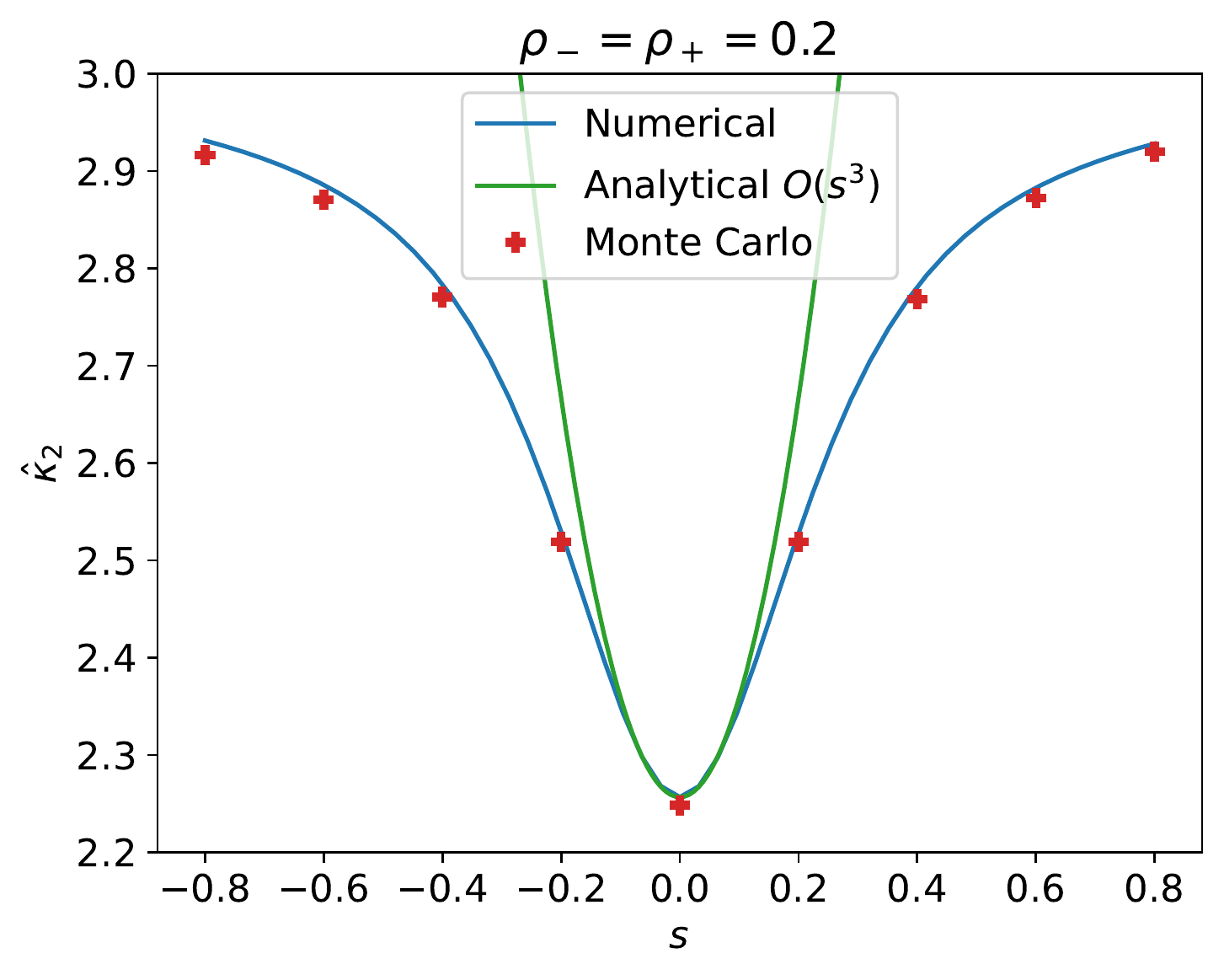} 
  \caption{Left: Profile $\Phi_1^{(2)}(v)$ at $\rho=0.6$~(\ref{eq:q12MFTBiais}) (solid blue), compared to the numerical resolution of the MFT equations \eqref{eq:MFT_qt} and  \eqref{eq:MFT_pt} (dashed red). Right: rescaled cumulant $\hat\kappa_2$ as a function of the bias $s$, obtained from the numerical resolution of the MFT equations \eqref{eq:MFT_qt} and  \eqref{eq:MFT_pt} (solid blue), compared to the small bias expansion~\eqref{eq:Kappa2SmallBiais} (solid green line). The points are obtained from Monte Carlo simulations (15.8 million simulations, final time 100000). Note that the correction in $s^2$ to $\hat{\kappa}_2$ is always positive, for all the values of the density $\rho$.}
  \label{fig:q12Biais}
\end{figure}

In addition to fully characterize the bath-tracer correlations, the generalized density profiles $\Phi_n$ also lead to the cumulants of the tracer's position. This is made possible by the key relation derived above [Eq. \eqref{eq:BoundPhi0_bis}]. We get, for $\hat \kappa_n \equiv \lim_{t\to\infty}[\kappa_n/\sqrt{2t}]$,
\begin{eqnarray}
  \hat \kappa_1 &=& s \frac{1-\rho}{\rho \sqrt{\pi}} + \mathcal{O}(s^2),
 \quad
  \hat \kappa_2 = \frac{1-\rho}{\rho \sqrt{\pi}}  + \mathcal{O}(s^2)
 , \\
  \hat \kappa_3 &=& \frac{s}{{\pi ^{3/2} \rho ^3}} \left[ (1-\rho ) \left(
      12 (1-\rho )^2
      -\pi  \left(\left(8-3 \sqrt{2}\right) \rho ^2\right.\right.\right.\nonumber\\
      &&\left.\left.\left. -3 \left(4-\sqrt{2}\right) \rho +3\right)
    \right) \right]
  + \mathcal{O}(s^2)
  \:.
\end{eqnarray}
We notice that, up to order $n=3$, $\hat \kappa_{n} = s \: \hat \kappa_{n+1}^{(s=0)} + \mathcal{O}(s^2)$, which implies that
\begin{equation}
\label{fluct_relation}
  \psi(\lambda,t) \underset{t\to\infty}{\sim} \psi^{(s=0)}(\lambda,t) + s \: \dt{\psi}{\lambda}^{(s=0)} + \mathcal{O}(s^2,\lambda^4)
  \:.
\end{equation}
On top of that, we checked from the high-density solution obtained in \cite{Illien2013a,Poncet2021a} that, when $\rho\to 1$, Eq. \eqref{fluct_relation} holds at any order in $\lambda$, and at arbitrary time. This points towards the generality of this relation.

\emph{Beyond linear response.---} We next show that explicit analytical results can be obtained beyond linear response which, as we proceed to show, can be quantitatively and even qualitatively significant. In addition, even if our previous expressions provide the leading order in the bias $s$, they do not bring non-trivial information for even cumulants, since the first non-zero correction to the unbiased case is actually of order $s^2$ for symmetry reasons. We thus compute the profile $\Phi_1$ at quadratic order in the bias, and get~\cite{SM}
\begin{widetext}
    \begin{multline}
  \label{eq:q12MFTBiais}
  \Phi_1^{(2)}(v) =
  \frac{(1-2 \rho) (1-\rho)^2}{2\rho ^2} \erfc(v)
  +\frac{(3-\rho)  (1-\rho)^2 }{\pi  \rho ^2} \erfc(v)
  -\frac{(1-\rho)^2 }{2\rho}\text{erfc}\left(\frac{v}{\sqrt{2}}\right)^2
  \\
  -\frac{(1-\rho)^2}{2\rho ^2} G(\sqrt{2}\: v)
  +\frac{5  (1-\rho)^3  }{\pi ^{3/2} \rho ^2}v \: \e^{-v^2}
  -\frac{(3-5 \rho ) (1-\rho)^2  }{\pi  \rho ^2} \e^{-v^2}
  -\frac{(1-\rho )^2 }{\pi ^{3/2} \rho ^2} v \: \e^{-\frac{v^2}{2}} \: \mathrm{K}_0\left(\frac{v^2}{2}\right)
   \:.
\end{multline}
\end{widetext}
Interestingly, we note that, even at order $1$ in $\lambda$ (and not only at order $2$ as in the linear response analysis discussed above), the density profile is in fact non-analytic at the origin. 
We stress that this qualitatively different feature emerges \textit{beyond} linear response.

In addition, the expression of $\Phi_1^{(2)}$ yields the $s^2$ order of $\hat\kappa_2 =  \left. \hat\kappa_2 \right|_{s=0} +s^2 \: \Delta \hat\kappa_2^{(2)} + \mathcal{O}(s^3)$, with
\begin{equation}
  \label{eq:Kappa2SmallBiais}
  \Delta \hat\kappa_2^{(2)} = 
  \frac{(1-\rho)^2 (7-5 \rho -\pi ((\sqrt{2}-3) \rho +2))}{\pi ^{3/2} \rho ^3} 
  \:.
\end{equation}
This result constitutes the first determination of the bias-dependence of the variance of a driven tracer in the SEP for an arbitrary density, a problem which has remained open for more than 25 years.

The function $ \Phi_1^{(2)}(v)$ is plotted in Fig.~\ref{fig:q21Biais} and displays very good agreement with the results obtained from the numerical procedure described above. We also display the dependence of the second cumulant as a function of the bias for a given value of the density $\rho=0.2$, which shows good agreement with both microscopic Monte Carlo simulations and the numerical resolution as long as the bias is small enough. This cumulant displays an important variation with the bias ($\sim 30 \%$), emphasising the quantitative importance of studying the problem beyond linear response (which gives zero variation).

\emph{Conclusion.---} In this Letter, starting from microscopic considerations, we built a hydrodynamic framework to study both the dynamics of a driven tracer in the SEP and the response of its environment. This allowed us to determine the first cumulants of bath-tracer correlations and of the tracer position at linear order in the bias and at arbitrary density -- a regime of parameters that was left aside so far. We also went beyond linear response by determining the second cumulant and the corresponding correlation profile, therefore unveiling for the first time the dependence of the variance of the tracer's position on the bias. Importantly, this approach is general and can be extended to study other models of single-file transport, by replacing in Eqs.~\eqref{eq:MFT_qt}-\eqref{eq:ContJ0} the transport coefficients $D$ and $\sigma$ by those of the system under consideration, and adapting the matching condition~\eqref{eq:biais_cond_dual} which can be derived from microscopic considerations, as done here for the SEP.

\emph{Acknowledgements.---} We thank Alexis Poncet for numerous discussions at early stages of this work, both on analytical and numerical aspects.


%

\end{document}


\setcounter{equation}{0}
\setcounter{figure}{0}
\setcounter{table}{0}
\setcounter{page}{1}
\makeatletter
\renewcommand{\theequation}{S\arabic{equation}}
\renewcommand{\thefigure}{S\arabic{figure}}
\renewcommand{\bibnumfmt}[1]{[S#1]}
\renewcommand{\citenumfont}[1]{S#1}

\title{Supplementary Material for Driven tracer in the Symmetric Exclusion Process: Linear Response and Beyond}

\author{Aur\'elien Grabsch}

\author{Pierre Rizkallah}

\author{Pierre Illien}

\author{Olivier B\'enichou}

\maketitle

\tableofcontents

\section{Microscopic equations and hydrodynamic limit}
\label{sec:sm_Micro}

We briefly recall the derivation of the microscopic equations presented in Refs.~\cite{Poncet:2021,Grabsch:2022}, but for the case of a biased tracer~\cite{ThesisPoncet}. Our starting point is the master equation of the SEP, which reads
\begin{align}
    \label{eq:sm_master}
    \partial_t P(X, \ueta, t) 
    =&~ \frac{1}{2}\sum_{r\neq X, X-1} \left[P(X, \ueta^{r,+}, t)-P(X, \ueta, t)\right] \nonumber 
    \\
    &+ \sum_{\mu=\pm 1} \frac{1+\mu s}{2} \left\{ (1 - \eta_X) P(X-\mu, \ueta, t) 
    - (1-\eta_{X+\mu}) P(X, \ueta, t) \right\} 
    \:,
\end{align}
where $P(X, \ueta, t)$ is the probability to observe at time $t$ the bath particles in a configuration described by the occupations of the sites $\ueta = \{ \eta_r = 0,1 \}$, and with the tracer at position $X$. The first term corresponds to the motion of the bath particles, while the second one describes the motion of the tracer, which can hop to the right with rate $\frac{1+s}{2}$ and to the left with rate $\frac{1-s}{2}$, with $s$ the bias on the dynamics of the tracer.

We consider the cumulant generating function of the position of the tracer
\begin{equation}
    \psi(\lambda,t) = \ln \moy{\e^{\lambda X_t}}
    \:.
\end{equation}
Its time evolution is deduced from the master equation~\eqref{eq:sm_master}, and reads~\cite{ThesisPoncet}
\begin{equation}
    \label{eq:sm_evolPsi}
    \partial_t \psi 
    =  \frac{1+s}{2}(\e^\lambda-1)(1-w_1) 
    +  \frac{1-s}{2} (\e^{-\lambda}-1)(1-w_{-1})
    \:,
\end{equation}
where we have denoted
\begin{equation}
    w_r(t) = \frac{\moy{\eta_{X_t+r}\e^{\lambda X_t}}}
    {\moy{\e^{\lambda X_t}}}
    \:.
\end{equation}
We call $w_r$ the generalised density profile generating function, since by expanding it in powers of $\lambda$ it generates all correlation functions between the displacement of the tracer and the density of bath particles at a distance $r$ from the tracer (represented by the occupation number $\eta_{X_t+r}$):
\begin{equation}
    w_r(t) = \sum_{n \geq 0} \frac{\lambda^n}{n!}
    \moy{\eta_{X_t+r} X_t^n}_c
    \:,
\end{equation}
with $\moy{\cdots}_c$ the joint cumulants. For instance, $\moy{\eta_{X_t+r} X_t}_c = \moy{\eta_{X_t+r} X_t} - \moy{\eta_{X_t+r}} \moy{ X_t}$.

Similarly, we can also write from the master equation the time evolution of $w_r$. We will only need the time evolution of the profiles just in front and behind the tracer~\cite{ThesisPoncet}:
\begin{equation}
    \label{eq:sm_evol_w_pm1}
    \partial_t w_{\pm 1} 
    = \frac{1}{2} \nabla_\pm w_{\pm 1} + 
    \frac{\partial_t \psi}{\e^{\pm \lambda}-1} w_{\pm 1} 
    +  \frac{1\pm s}{2} \e^{\pm\lambda} f_{\pm 1, \pm 2} -  \frac{1\mp s}{2} f_{\mp 1, \pm 1}
    \:,
\end{equation}
where we have introduced the higher order correlation functions
\begin{equation}
    \label{eq:sm_def_f}
    f_{\mu, r}(\lambda, t) \equiv \displaystyle
    \frac{\left\langle(1-\eta_{X_t+\mu})\eta_{X_t+r}\e^{\lambda X_t}\right\rangle}{\langle \e^{\lambda X_t}\rangle} - 
    \begin{cases}
        (1-w_\mu) w_{r-\mu} & \text{ if } \mu r > 0 \:, \\
        (1-w_\mu) w_r  & \text{ if } \mu r < 0 \:.
    \end{cases}
\end{equation}

In the hydrodynamic limit of large time and large distances, the different observables have the scaling~\cite{Poncet:2021,Grabsch:2022},
\begin{equation}
    \label{eq:sm_hydro_scalings}
    \psi(\lambda,t) \underset{t \to \infty}{\simeq}
    \hat{\psi}(\lambda) \: \sqrt{2t}
    \:,
    \quad
    w_r(t) \underset{t \to \infty}{\simeq}
    \Phi \left(
    v = \frac{r}{\sqrt{2t}}
    \right)
    \:,
    \quad
    f_{\mu,r} \underset{t \to \infty}{\simeq}
    \frac{1}{\sqrt{2t}} F_\mu \left(
    v = \frac{r}{\sqrt{2t}}
    \right)
    \:,
\end{equation}
where we have omitted the dependency on $\lambda$ of $\Phi$ and $F_\mu$ for simplicity. Plugging these scaling forms into the evolution equations~(\ref{eq:sm_evolPsi},\ref{eq:sm_evol_w_pm1}), we obtain
\begin{equation}
    \label{eq:sm_BoundPhi0}
    (1+s) (\e^{\lambda}-1) (1-\Phi(0^+)) 
    + (1-s) (\e^{\lambda}-1)(1-\Phi(0^-))
    = 0
    \:,
    \quad
    \Phi'(0^\pm) \pm \frac{2 \hat{\psi}}{\e^{\pm \lambda}-1}
    \Phi(0^\pm)
    = 0
    \:.
\end{equation}
Remarkably, in the evolution equation for $w_{\pm 1}$~\eqref{eq:sm_evol_w_pm1}, the contributions of the functions $F_\mu$, which involve higher order correlations, vanish. This would not be the case for general $w_r$ with $r \neq 1$, as this would yield an equation which is not closed. In the unbiased case, where the tracer has the same dynamics as the other particles, a closure relation was found that allows the determination of $\Phi$~\cite{Grabsch:2022}. Here, such closure is not available, therefore we will rely on another approach.

\section{Macroscopic fluctuation theory for the profiles}
\label{sec:sm_MFT}

At large scale and large distances, the SEP (with no biased tracer) can by described in the framework of fluctuating hydrodynamics~\cite{Spohn:1991}. The bath particles are described by a continuous density $\rho(x,t)$, which satisfies a conservation relation
\begin{equation}
    \label{eq:sm_ConsRho}
    \partial_t \rho + \partial_x j = 0
    \:,
\end{equation}
with a stochastic current $j$, assumed to be Gaussian, with,
\begin{equation}
    \label{eq:sm_StochCurrent}
    j(x,t) = - D(\rho(x,t)) \partial_x \rho(x,t)
    + \sqrt{\sigma(\rho(x,t))} \: \eta(x,t)
    \:,
\end{equation}
with $\eta$ a Gaussian white noise, uncorrelated in space and time,
\begin{equation}
    \moy{\eta(x,t) \eta(x',t')} = \delta(x-x') \delta(t-t')
    \:.
\end{equation}
In this formalism, all the microscopic details of the system are encoded in the two transport coefficients, which, for the SEP, read~\cite{Bodineau:2004,Derrida:2009a}
\begin{equation}
    \label{eq:sm_tr_coefs_SEP}
    D(\rho) = \frac{1}{2}
    \:,
    \qquad
    \sigma(\rho) = \rho(1-\rho)
    \:.
\end{equation}

This formalism is however difficult to manipulate in practice due to the Gaussian white noise $\eta(x,t)$. A more convenient formulation of fluctuating hydrodynamics, which is deterministic, is available: the Macroscopic Fluctuation Theory (MFT). For details on the construction of the MFT from the equations of fluctuating hydrodynamics~(\ref{eq:sm_ConsRho},\ref{eq:sm_StochCurrent}), see Ref.~\cite{Derrida:2009a} (it is also reproduced in the Appendix of Ref.~\cite{Rizkallah:2022}). The MFT gives the probability to observe an evolution $\rho(x,t)$ of the density, between time $t=0$ and $t=T$, as a functional integral,
\begin{equation}
    \label{eq:sm_MFT_action}
    P[\{ \rho(x,t) \}] \propto \int \mathcal{D} H \:
    \e^{- S[\rho,H]}
    \:,
    \quad
    S[\rho,H] = 
    \int_{0}^T \dd t \int_{-\infty}^\infty \dd x \: \left[
    H \partial_t \rho + D(\rho) \partial_x \rho \partial_x H
    - \frac{\sigma(\rho)}{2} (\partial_x H)^2
    \right]
    \:.
\end{equation}
The field $H(x,t)$ is a conjugate field that has been introduced to handle the constraint~\eqref{eq:sm_ConsRho} on $j$ and $\rho$. It is related to the current $j$ by
\begin{equation}
    j = - D(\rho) \partial_x \rho + \sigma(\rho) \partial_x H
    \:.
\end{equation}
At $t=0$, the initial density profile $\rho(x,0)$ is picked from the equilibrium distribution of the SEP, which reads
\begin{equation}
    \label{eq:sm_InitWeight_MFT}
    P[\{ \rho(x,0) \}] \propto \e^{-F[\rho(x,0)]}
    \:,
    \quad
    F[\rho(x,0)] = \int \dd x \int_{\rho}^{\rho(x,0)} \dd z \: 
  \frac{2 D(z)}{\sigma(z)}( \rho(x,0)-z)
  \:,
\end{equation}
where $\rho$ is the probability that each site is initially occupied.
In this formalism, the position of the tracer can be obtained from the density $\rho(x,t)$ by writing that the number of particles on the right of the tracer is conserved~\cite{Krapivsky:2015a}:
\begin{equation}
    \label{eq:sm_Xt_from_rho}
     \int_0^{X_t[\rho]} \rho(x,t) \dd x = \int_0^\infty (\rho(x,t) - \rho(x,0))\dd x
    \:.
\end{equation}
This allows to obtain the statistical properties of observables by computing their moment generating functions, for instance,
\begin{equation}
    \label{eq:sm_MFT_MomentsXt}
    \moy{\e^{\lambda X_T}}
    = \int \mathcal{D}\rho(x,0)
    \int \mathcal{D}\rho(x,t) \mathcal{D}H(x,t)
    \e^{-S[\rho,H] - F[\rho(x,0)] + \lambda X_T}
    \:.
\end{equation}
One can rescale the time by $T$ and the positions by $\sqrt{T}$ to extract the $T$ dependence of the fields. This yields that the argument of the exponential is proportional to $\sqrt{T}$. Hence for large $T$, the integral in~\eqref{eq:sm_MFT_MomentsXt} is dominated by a single configuration $(q,p)$ of $(\rho,H)$, obtained by minimising the action. This yields the MFT equations~\cite{Derrida:2009a}
\begin{equation}
    \partial_t q 
    =
    \partial_x (D(q)\partial_x q)
    - \partial_x (\sigma(q) \partial_x p)
    \:,
    \quad
     \partial_t p
    = -D(q) \partial_x^2 p
    - \frac{\sigma'(q)}{2} (\partial_x p)^2
    \:,
\end{equation}
with the terminal condition on $p$~\cite{Krapivsky:2015a},
\begin{equation}
    \label{eq:sm_MFT_termp}
    p(x,T) = \frac{\lambda}{q(Y,T)} \Theta(x-Y)
    \:,
    \quad
    Y = X_T[q]
    \:,
\end{equation}
and initial condition on $q$,
\begin{equation}
  \label{eq:sm_MFT_initq}
  p(x,0) = \frac{\lambda}{q(Y,T)} \Theta(x) + \int_{\rho}^{q(x,0)} \dd r \frac{2 D(r)}{\sigma(r)}
  \:.
\end{equation}
These equations have been solved perturbatively in~\cite{Krapivsky:2015a} for the lowest orders in $\lambda$. One difficulty is that the final profile $q(x,T)$ is discontinuous at $x=Y$, and so one must find a way to make sense of $q(Y,T)$. Here, we will circumvent this difficulty by using a mapping onto a different problem.

\subsection{Computing the profiles from MFT}
\label{sec:sm_MFT_prof}

The formalism of MFT, which we will adapt to describe a biased tracer, can be used to compute the profile $\Phi(v)$ which is one of our main observables. Indeed, we can write
\begin{align}
    w_r(T) 
    &= \frac{\moy{\eta_{X_T+r} \e^{\lambda X_T}}}{\moy{\e^{\lambda X_T}}}
    = \frac{\displaystyle
    \int \mathcal{D}\rho(x,0)
    \int \mathcal{D}\rho(x,t) \mathcal{D}H(x,t) 
    \: \rho(X_T[\rho] + r,T) \:
    \e^{-S[\rho,H] - F[\rho(x,0)] + \lambda X_T}}
    {\displaystyle \int \mathcal{D}\rho(x,0)
    \int \mathcal{D}\rho(x,t) \mathcal{D}H(x,t)
    \e^{-S[\rho,H] - F[\rho(x,0)] + \lambda X_T}}
    \nonumber
    \\
    &\underset{T \to \infty}{\simeq}
    q \left( X_T + r \sqrt{T} , T \right)
    \:,
\end{align}
by again evaluating the integrals with a saddle point method (the saddle point is identical for the numerator and the denominator).
Therefore, from the scaling~\eqref{eq:sm_hydro_scalings},
\begin{equation}
    \label{eq:sm_Phi_from_q}
    \Phi(v) = q( X_T + v \sqrt{2}, 1 )
    \:,
\end{equation}
where we have set $T=1$ for simplicity. The MFT profile at final time $q(x,1)$ thus coincides with the large time scaling function $\Phi$ of our generalised density profile.

We have here reproduced the argument presented in Refs.~\cite{Poncet:2021,Grabsch:2022} to show that $\Phi$ can be computed from MFT. This is still the case for a biased tracer, as we will see that the effect of a bias can be taken into account by implementing specific boundary conditions at $X_t$ for both $\rho$ and $H$. This does not change the argument presented here.

\subsection{Bias matching condition}

Our goal is to find the matching condition that must be implemented on the density $\rho(x,t)$ at $x=X_t$ in order to take into account the biased dynamics of the tracer. This can be done by inspecting the boundary condition that we obtained microscopically~\eqref{eq:sm_BoundPhi0}, which combined with~\eqref{eq:sm_Phi_from_q}, yields the naive relation
\begin{equation}
    \frac{1+s}{1-s} \frac{1 - q(X_t^+,t)}{1-q(X_t^-,t)}
    = \e^{-\lambda}
    \:.
\end{equation}
However, this relation cannot be correct since for an unbiased tracer ($s=0$), this would imply that $q(x,t)$ is discontinuous at $X_t$ for all times $t$. However, it is well-known that, for an unbiased tracer, $q(x,t)$ is only discontinuous at the final time $t=1$~\cite{Derrida:2009a,Krapivsky:2015a} (and at initial time $t=0$). Therefore, we must remove this artificial discontinuity due to $\lambda$, which only arises because $\Phi$ corresponds to the final profile due to~\eqref{eq:sm_Phi_from_q}. By doing so, we obtain the bias condition
\begin{equation}
    \label{eq:sm_bias_cond_opt}
    (1+s) (1-q(X_t^+,t) ) = (1-s) (1-q(X_t^-,t))
    \:.
\end{equation}
We have written this relation for the optimal profile $q$ only, which depends on the parameter $\lambda$. However, relation~\eqref{eq:sm_bias_cond_opt} does not involve explicitly $\lambda$. Furthermore, in the absence of bias ($s=0$) all realisations of the fluctuating field $\rho(x,t)$ are continuous. The addition of the biased dynamics of the tracer is expected to create a discontinuity for all realisations of the stochastic profile $\rho(x,t)$ (due to the accumulation of particles in front of the tracer). Since the external force applied on the tracer does not depend on the density of surrounding particles, we assume that the acts in the same way for all the realisations, hence,
\begin{equation}
    \label{eq:sm_bias_cond}
    (1+s) (1-\rho(X_t^+,t) ) = (1-s) (1-\rho(X_t^-,t))
    \:.
\end{equation}
This extends the relation written in~\cite{Landim:1998} for $\lambda=0$ to arbitrary value of $\lambda$. A similar relation was written in~\cite{Kundu:2016} for a different model.

\subsection{Mapping onto the dual flux problem}

Implementing the bias boundary condition~\eqref{eq:sm_bias_cond} is difficult because the position $X_t$ is not fixed. Additionally, the MFT for a tracer is rather cumbersome due to the ill-defined factor $q(Y,1)$ that appears in the boundary conditions~(\ref{eq:sm_MFT_termp},\ref{eq:sm_MFT_initq}). We can actually solve the two problems at once by using a well-known duality relation that maps the SEP onto another model: the zero-range process~\cite{Evans:2000,Evans:2005}. This duality was  recently discussed in  details for general systems (not only the SEP) in~\cite{Rizkallah:2022}. The mapping is more easily written  at the level of the stochastic fields $\rho(x,t)$ and $j(x,t)$ of the fluctuating hydrodynamics. We define the dual fields $\rt$ and $\jt$ as 
\begin{equation}
    \rho(x,t) = \frac{1}{\rt(k(x,t),t)}
    \:,
    \quad
    j(x,t) = -\frac{\jt(k(x,t),t)}{\rt(k(x,t),t)}
    \:,
    \quad
    k(x,t) = \int_0^x \rho(x',t) \dd x' - \int_0^t j(0,t') \dd t'
\end{equation}
This mapping actually relies on the fact that one can equivalently describe the gaps between the particles in the SEP (dual fields $\rt$ and $\jt$) or the positions of the particles (original fields $\rho$ and $j$)~\cite{Rizkallah:2022}. These new fields still obey the equations of fluctuating hydrodynamics~(\ref{eq:sm_ConsRho},\ref{eq:sm_StochCurrent}), but with $D$ and $\sigma$ replaced respectively by
\begin{equation}
    \label{eq:sm_dual_transp_coefs}
    \tilde{D}(\rt )
    = \frac{1}{\rt^2} 
    D \left( \frac{1}{\rt} \right)
    = \frac{1}{2 \rt^2}
    \:,
    \qquad
     \tilde{\sigma}(\rt )
    = \rt \:
    \sigma \left( \frac{1}{\rt} \right)
    = 1 - \frac{1}{\rt}
    \:.
\end{equation}

The inverse transformation is identical, and given by
\begin{equation}
    \label{eq:sm_Inverse_Mapping}
    \rt(k,t) = \frac{1}{\rho(x(k,t),t)}
    \:,
    \quad
    \jt(k,t) = -\frac{j(k(x,t),t)}{\rho(k(x,t),t)}
    \:,
    \quad
    x(k,t) = \int_0^k \rt(k',t)\dd k' - \int_0^t \jt(0,t') \dd t'
    \:.
\end{equation}
Under this duality, the position $X_t$ of the tracer, defined as~\eqref{eq:sm_Xt_from_rho}, becomes
\begin{equation}
    \label{eq:sm_Xt_from_Qt}
    X_t
    = -\int_0^\infty (\rt(k,t) - \rt(k,0)) \dd k
    \equiv
    - \tilde{Q}_t
    \:,
\end{equation}
where $\tilde{Q}_t$ is the integrated current through the origin in the dual system,
\begin{equation}
    \tilde{Q}_t = \int_0^t \jt(0^+,t') \dd t'
    \:,
\end{equation}
which is equivalent to~\eqref{eq:sm_Xt_from_Qt} due to the conservation relation~\eqref{eq:sm_ConsRho}.

Note that we would have the same relation with $j(0^-,t)$ if we had used the equivalent definition
\begin{equation}
  \tilde{Q}_t = - \int_{-\infty}^0 (\rt(k,t) - \rt(k,0)) \dd k
  = \int_0^t \jt(0^-,t') \dd t'
  \:.
\end{equation}
This implies that
\begin{equation}
    \label{eq:sm_cont_j}
    \dt{\tilde{Q}_t}{t} = \jt(0^-,t) = \jt(0^+,t)
    \:,
\end{equation}
hence the current $\jt$ must be continuous at the origin in order to have a coherent definition of $\tilde{Q}_t$ on both sides. With these relations, the mapping of the positions~\eqref{eq:sm_Inverse_Mapping} can be written as
\begin{equation}
    \label{eq:sm_tr_pos_mapping}
  x(k,t) = X_t + \int_0^k \rt(k',t) \dd k'
  \:.
\end{equation}
It is thus straightforward to see that (since the densities are positive):
\begin{equation}
  \left\lbrace
    \begin{array}{l}
      x(0^+,t) = X_t^+
      \\
      x(0^-,t) = X_t^-
    \end{array}
  \right.
  \quad \Leftrightarrow \quad
    \left\lbrace
    \begin{array}{l}
      k(X_t^+,t) = 0^+
      \\
      k(X_t^-,t) = 0^-
    \end{array}
  \right.
\end{equation}
Consequently, with this mapping, we replace the moving position $X_t$
with the fixed origin. This gets rid of the difficulty of tracking the tracer, and yields the new boundary condition for the dual system
\begin{equation}
    \label{eq:sm_biais_cond_dual}
    (1+s)\left( 1 - \frac{1}{\rt(0^+,t)} \right)
    = (1-s) \left( 1 - \frac{1}{\rt(0^-,t)} \right)
    \:.
\end{equation}

\subsection{MFT equations for the dual flux problem}

Having mapped the problem of a biased tracer onto a dual flux problem at the level of the fluctuating hydrodynamics, we now switch to a MFT description of the field $\rt$ and a new conjugate field $\tilde{H}$ related to $\jt$ by 
\begin{equation}
    \label{eq:sm_current_dual_MFT}
    \jt = - \tilde{D}(\rt) \partial_k \rt + \tilde{\sigma}(\rt) \partial_k \Ht
    \:.
\end{equation}
These fields obey the MFT action~\eqref{eq:sm_MFT_action}, but for the dual system,
\begin{equation}
    \tilde{S}[\rt,\Ht] = 
    \int_{0}^1 \dd t \int_{-\infty}^\infty \dd k \: \left[
    \Ht \partial_t \rt + \Dt(\rt) \partial_k \rt \partial_k \Ht
    - \frac{\st(\rt)}{2} (\partial_k \Ht)^2
    \right]
    \:,
\end{equation}
and the initial condition given by a weight $\e^{-\tilde{F}[\rt(x,0)]}$ as in~\eqref{eq:sm_InitWeight_MFT}, with
\begin{equation}
    \tilde{F}[\rt(x,0)] = \int \dd k \int_{\rt}^{\rt(k,0)} \dd z \: 
  \frac{2 \Dt(z)}{\st(z)}( \rt(k,0)-z)
  \:,
\end{equation}
where $\rt = 1/\rho$ is the mean density of the dual system.
Using that the position $X_t$ of the tracer is directly related to the current in the dual model~\eqref{eq:sm_Xt_from_Qt}, we can write the moment generating function as
\begin{equation}
    \label{eq:sm_MFT_MomentsXtfromQt}
    \moy{\e^{\lambda X_T}}
    = \moy{\e^{-\lambda \tilde{Q}_T}}
    = \int \mathcal{D}\rt(k,0)
    \int \mathcal{D}\rt(k,t) \mathcal{D} \Ht(k,t)
    \e^{-\tilde{\mathscr{S}}[\rt,\Ht]}
    \:,
    \quad
    \tilde{\mathscr{S}}[\rt,\Ht] =
    \tilde{S}[\rt,\Ht] + \tilde{F}[\rt(k,0)] + \lambda Q_T[\rt]
\end{equation}
with
\begin{equation}
    Q_T[\rt] = \int_0^\infty \left(
        \rt(k,1) - \rt(k,0)
    \right) \dd k
    \:,
\end{equation}
where we have again set $T=1$ for convenience. This writing of the cumulant generating function of $X_T$ is similar to the one presented in Section~\ref{sec:sm_MFT} above, but here it is written in terms of the fields $\rt$ and $\Ht$ of the dual system, instead of the original fields $\rho$ and $H$. As argued in Section~\ref{sec:sm_MFT}, the integral over the fields is dominated by the optimal configuration $(\qt,\pt)$ which minimises the action $\tilde{\mathscr{S}}[\rt,\Ht]$. The equations to determine this minimum can be obtained by computing the variation of the action,
\begin{align}
    \tilde{\mathscr{S}}[\qt + \delta \rt,\pt + \delta \Ht] 
    &= \tilde{\mathscr{S}}[\rt,\Ht]
    + \int_0^1 \dd t \int_{-\infty}^{+\infty} \dd k
    \Bigg\lbrace
    \delta \rt(k,t) \left[
    - \partial_t \pt - \Dt(\qt) \partial_k^2 \pt
    - \frac{\st'(\qt)}{2} (\partial_k \pt)^2
    \right]
    \nonumber
    \\
    &+ \delta \Ht(k,t) \left[
    \partial_t \qt - \partial_k(\Dt(\qt) \partial_k \qt)
    + \partial_k (\st(\qt) \partial_k \pt)
    \right]
    \Bigg\rbrace
    \nonumber
    \\
    &+ \int_{-\infty}^{+\infty} \dd k
    \Bigg\lbrace \delta \rt(k,1) \left[ \pt(k,1)
    +\lambda \Theta(k)
    \right]
    + \delta \rt(k,0) \left[ -\pt(k,0)
    + \int_{\rt}^{\rt(k,0)} \frac{2 \Dt(z)}{\st(z)} \dd z 
    - \lambda \Theta(k)
    \right]
    \Bigg\rbrace
    \nonumber
    \\
    &+ \int_0^1 \dd t \Bigg\lbrace
    \left[ -\delta \rt \Dt(\qt) \partial_k \pt  \right]_{0^-}^{0^+}
    + 
    \left[ \delta \Ht (-\Dt(\qt) \partial_k \qt + \st(\qt) \partial_k \pt)  \right]_{0^-}^{0^+}
    \Bigg\rbrace
        \label{eq:sm_Min_action}
    \:,
\end{align}
where $\Theta$ is the Heaviside step function, $[f]_{0^-}^{0^+} = f(x=0^+)-f(x=0^-)$, and we have used integration by parts while taking into account the fact that the fields $\qt$, $\pt$ and their derivatives can be discontinuous at $0$ due to the presence of the bias.  Imposing that the different terms vanish, we get from the first two lines the MFT equations, valid for $k>0$ and $k<0$,
\begin{equation}
    \label{eq:sm_MFT_bulk_dual}
    \partial_t \qt 
    = \partial_k(\Dt(\qt) \partial_k \qt)
    - \partial_k (\st(\qt) \partial_k \pt)
    \:,
    \qquad
    \partial_t \pt 
    = -\Dt(\qt) \partial_k^2 \pt
    - \frac{1}{2}\st'(\qt) (\partial_k \pt)^2
    \:.
\end{equation}
The second line of~\eqref{eq:sm_Min_action} gives both the initial and final conditions
\begin{equation}
    \label{eq:sm_bound_pt}
    \pt(k,0) = 
    \int_{\rt}^{\qt(k,0)} \frac{2 \Dt(z)}{\st(z)} \dd z 
    - \lambda \Theta(k)
    \:,
    \quad
    \pt(k,1)
    = -\lambda \Theta(k)
    \:,
\end{equation}
Up to now, these are the usual MFT equations for the problem of the flux through the origin~\cite{Derrida:2009a}. However, the last line of~\eqref{eq:sm_Min_action} gives two new equations, specific to the presence of a bias,
\begin{equation}
     \label{eq:sm_disc_from_action}
     \left[ -\delta \rt \Dt(\qt) \partial_k \pt  \right]_{0^-}^{0^+} = 0
     \:,
     \qquad
     \left[ \delta \Ht (-\Dt(\qt) \partial_k \qt + \st(\qt) \partial_k \pt)  \right]_{0^-}^{0^+}
     =  \left[ \delta \Ht \jt  \right]_{0^-}^{0^+}
     \:,
\end{equation}
where we have used the relation with the current~\eqref{eq:sm_current_dual_MFT}. Since we have seen in Eq.~\eqref{eq:sm_cont_j} that the current must be continuous at the origin, this imposes
\begin{equation}
     \delta \Ht(0^+,t) =  \delta \Ht(0^-,t)
     \:.
\end{equation}
We thus choose $\Ht(0^+,t) = \Ht(0^-,t)$ to enforce this condition, which becomes for the typical realisation
\begin{equation}
    \label{eq:sm_cont_pt}
     \pt(0^+,t) = \pt(0^-,t)
     \:.
\end{equation}
The variation of the bias condition~\eqref{eq:sm_biais_cond_dual} imposes
\begin{equation}
    (1+s)  \frac{\delta\rt(0^+,t)}{\qt(0^+,t)^2 }
    = (1-s) \frac{\delta \rt(0^-,t)}{\qt(0^-,t)^2 }
    \:,
\end{equation}
which, combined with the first equation in~\eqref{eq:sm_disc_from_action} yields a relation for the derivative of $\pt$,
\begin{equation}
    \label{eq:sm_disct_dpt}
    (1-s) \partial_k \pt(0^+,t)
    = (1+s) \partial_k \pt(0^-,t)
    \:.
\end{equation}
The MFT equations~\eqref{eq:sm_MFT_bulk_dual}, together with the initial and terminal boundary conditions~\eqref{eq:sm_bound_pt} and the relations at the origin for $\qt$, $\pt$ and their derivatives~(\ref{eq:sm_cont_j},\ref{eq:sm_biais_cond_dual},\ref{eq:sm_cont_pt},\ref{eq:sm_disct_dpt}) fully determine the solutions $\qt(k,t)$ and $\pt(k,t)$. These equations thus allow to compute the profiles $\Phi$ from relation~\eqref{eq:sm_Phi_from_q}, combined with the duality relation between the fields $q$ and $\qt$~(\ref{eq:sm_Inverse_Mapping},\ref{eq:sm_tr_pos_mapping}). Explicitly, this gives a parametric representation, which only requires the dual profile at final time,
\begin{equation}
    \label{eq:sm_Phi_from_qt}
    \Phi\left(v = \frac{y(k,1)}{\sqrt{2}} \right)
    = q(X_T + y(k,1),1) = \frac{1}{\qt(k,1)}
    \:,
    \quad
    y(k,t) = \int_0^k \qt(k',t) \dd k'
    \:.
\end{equation}

\section{Numerical resolution of the MFT equations}

At order 0 in $\lambda$, the MFT equations~\eqref{eq:sm_MFT_bulk_dual} simplify into a unique diffusion equation obeyed by the zeroth order $\qt_0$ of $\qt$,
\begin{equation}
    \partial_t \qt_0
    = \partial_k(\Dt(\qt_0) \partial_k \qt_0)
    \:.
\end{equation}
We solve this equation by a finite difference method associated to an explicit Euler method in time. We compute at each step the values at $t + \Delta t$ for all $x \neq 0^\pm$ using the discretized bulk diffusion equation. To ensure stability we choose the time step $\Delta t$ and space step $\Delta x$ such that $2 D \Delta t \leq \Delta x ^2 $ where $D = \max_{x,t} D(\tilde{q}_0(x,t))$ (evaluated heuristically). Then we compute the values at $0^\pm$ using the matching conditions~\eqref{eq:sm_cont_j} and~\eqref{eq:sm_biais_cond_dual}. Once $\qt_0$ is known, we can compute the first order of $\pt$ by the same method, now with the conditions~\eqref{eq:sm_cont_pt} and~\eqref{eq:sm_disct_dpt}. Then, higher-order terms follow similarly.
The MFT equations are singular near $t = 1$ (see the behavior near 0 on Fig.~2 in the main text), so numerically we have to restrict the resolution to one time step before $t = 1$.

\section{Perturbative resolution at lowest orders}

Even in the absence of a bias, the MFT equations are difficult to solve, or even impossible in most cases. Here, the main difficulty relies on the fact that the MFT equations~\eqref{eq:sm_MFT_bulk_dual} involve a non-constant diffusion coefficient. We can get rid of this difficulty by performing the transformation
\begin{equation}
  \hat{q}(x,t) = \frac{1}{\tilde{q}(k,t)}
  \quad \text{with} \quad
  x(k,t) = \int_0^k \tilde{q}(k',t) \dd k'
  \:,
\end{equation}
which is the inverse transformation~(\ref{eq:sm_Inverse_Mapping},\ref{eq:sm_tr_pos_mapping}) but in the reference frame of the tracer. The main advantage of proceeding  this way instead of trying to apply directly the MFT for the tracer is that we avoid the difficulty coming from the ill-defined value $q(X_t,t)$ which arises in this case (see Eq.~\eqref{eq:sm_MFT_termp}) for instance. We define a new conjugate field
\begin{equation}
    \hat{p}(x,t) = \pt(k,t)
    \:.
\end{equation}
With these definitions, the profile $\Phi$ can be straightforwardly deduced from~\eqref{eq:sm_Phi_from_qt}, which becomes
\begin{equation}
    \label{eq:sm_Phi_from_qh}
    \Phi(v) = \hat{q}(x=v \sqrt{2},1)
    \:.
\end{equation}

We map the MFT equations~(\ref{eq:sm_MFT_bulk_dual},\ref{eq:sm_bound_pt}) and the boundary conditions~(\ref{eq:sm_cont_j},\ref{eq:sm_biais_cond_dual},\ref{eq:sm_cont_pt},\ref{eq:sm_disct_dpt}) on these new fields by using
\begin{equation}
    \partial_k = \dep{x}{k} \partial_x = \frac{1}{\hat{q}} \partial_x
    \:,
    \quad
    \partial_t = \dep{x}{t} \partial_x + \partial_t
    \:,
\end{equation}
with
\begin{equation}
     \dep{x}{t} = \int_0^k \partial_t \qt(k',t) \dd k'
    = - \jt(k,t) + \jt(0,t)
    = \left[ \frac{1}{\hat{q}}D(\hat{q}) \partial_x \hat{q}
        + \frac{1}{\hat{q}^2} \sigma(\hat{q}) \partial_x \hat{p}
    \right]_0^{t}
    \:,
\end{equation}
with $D$ and $\sigma$ given by~\eqref{eq:sm_tr_coefs_SEP}. This gives for instance the equations
\begin{align}
    \label{eq:sm_MFT_mapped_back}
    \partial_t \hat{q} 
    - \dt{X_t}{t} \partial_x \hat{q}
    & =
    \partial_x (D(\hat{q})\partial_x \hat{q})
    - \partial_x \left(\frac{\sigma(\hat{q})}{\hat{q}} \partial_x \hat{p} \right)
    \:, \\
     \partial_t \hat{p}
    - \dt{X_t}{t} \partial_x \hat{p}
    & = -D(\hat{q}) \partial_x^2 \hat{p}
    - \frac{\sigma(\hat{q}) + \hat{q}\sigma'(\hat{q})}{2\hat{q}^2} (\partial_x \hat{p})^2
    + 2\dfrac{D(\hat{q})}{\hat{q}}(\partial_x \hat{p})(\partial_x \hat{q})
    \:,
\end{align}
with
\begin{equation}
    \dt{X_t}{t} = \left.
    \frac{1}{\hat{q}}D(\hat{q}) \partial_x \hat{q}
        + \frac{1}{\hat{q}^2} \sigma(\hat{q}) \partial_x \hat{p}
        \right|_{x=0}
        \:.
\end{equation}
These equations now involve the constant diffusion coefficient~\eqref{eq:sm_tr_coefs_SEP}, but still cannot be solved explicitly. 
We thus look for a perturbative solution in $\lambda$,
\begin{equation}
  \hat{q} = \hat{q}_0 + \lambda \hat{q}_1 + \lambda^2 \hat{q}_2 + \cdots
  \:,
  \quad
  \hat{p} = \lambda \hat{p}_1 + \lambda^2 \hat{p}_2 + \cdots
  \:.
\end{equation}

\subsection{Expressions at order \texorpdfstring{$0$}{0} in \texorpdfstring{$\lambda$}{lambda}}

At order $0$ in $\lambda$, our equations~\eqref{eq:sm_MFT_mapped_back}, along with the boundary conditions deduced from~(\ref{eq:sm_cont_j},\ref{eq:sm_biais_cond_dual}) coincide with those written and solved in~\cite{Landim:1998}. The solution assumes a  scaling form,
\begin{equation}
    \hat{q}_0(x,t) = \hat{q}_0 \left( \frac{x}{\sqrt{t}} \right)
    \:.
\end{equation}
Plugging this form into the equations allows for an explicit solution. Since, in the following, we will only need the expansion of $\hat{q}_0(x,t)$ for small $s$, we only give the first orders,
\begin{equation}
   \hat{q}_0(x>0,t) =
   \rho + s (1-\rho) \erfc\left(\frac{x}{\sqrt{2t}}\right)
   +\frac{(1-\rho)^2 s^2}{\rho}\left(
   \erfc\left(\frac{x}{\sqrt{2t}}\right)-\frac{2
   \e^{-\frac{x^2}{2 t}}}{\pi }\right)
   +O\left(s^3\right)
   \:.
\end{equation}
The solution for $x<0$ is deduced by the replacement $s \to -s$ and $x \to -x$.
This coincides with the small $s$ expansion of the result of~\cite{Landim:1998}, as it should.

\subsection{Expressions at order \texorpdfstring{$1$}{1} in \texorpdfstring{$\lambda$}{lambda}}

We now proceed similarly at first order in $\lambda$. The equations~\eqref{eq:sm_MFT_mapped_back} cannot be solved analytically for any bias $s$. We thus consider an expansion at small $s$,
\begin{equation}
  \hat{q}_1 = \hat{q}_1^{(0)} + s \hat{q}_1^{(1)} + s^2 \hat{q}_1^{(2)} + \cdots
  \:,
  \quad
  \hat{p}_1 = \hat{p}_1^{(0)} + s \hat{p}_1^{(1)} + s^2 \hat{p}_1^{(2)} + \cdots
\end{equation}
Solving~\eqref{eq:sm_MFT_mapped_back} with the boundary conditions~(\ref{eq:sm_cont_j},\ref{eq:sm_biais_cond_dual},\ref{eq:sm_cont_pt},\ref{eq:sm_disct_dpt}) yields, at order $0$,
\begin{equation}
    \hat{p}_1^{(0)}(x,t) = -\frac{1}{2}
    \erfc \left(
    - \frac{x}{\sqrt{2(1-t)}}
    \right)
    \:,
    \quad
    \hat{q}_1^{(0)}(x,t) =
    \frac{1-\rho}{2} \left( 
        \erfc \left(
    - \frac{x}{\sqrt{2(1-t)}}
        \right)
    -\erfc \left(
    - \frac{x}{\sqrt{2t}}
    \right)
    \right)
    \:.
\end{equation}

\subsubsection{Expressions at order \texorpdfstring{$1$}{1} in \texorpdfstring{$s$}{s}}

The MFT equations~\eqref{eq:sm_MFT_mapped_back} with the boundary conditions deduced from~(\ref{eq:sm_cont_j},\ref{eq:sm_biais_cond_dual},\ref{eq:sm_cont_pt},\ref{eq:sm_disct_dpt}) can be solved at order $1$ in $s$, and the solution reads,
\begin{multline}
    \hat{p}_1^{(1)}(x,t) =
    \frac{(1-\rho )}{2\rho} \left(4
   T\left(\frac{x}{\sqrt{t}},\sqrt{\frac{t}{1-t}}\right)+\text{erf}\left(\frac{x}{\sqrt{2-2 t}}\right) \text{erf}\left(\frac{x}{\sqrt{2}
   \sqrt{t}}\right)-\frac{2 \left(\sqrt{t}-1\right) e^{-\frac{x^2}{2
   (1-t)}}}{\pi  \sqrt{1-t}}+1\right)
   \\
   +
   \frac{1}{2\rho} \text{erfc}\left(\frac{x}{\sqrt{2} \sqrt{1-t}}\right)
   \:,
\end{multline}
\begin{multline}
    \hat{q}_1^{(1)}(x,t) =
    \frac{(1-\rho)}{2\rho } \left(-4 (1-\rho)
   T\left(\frac{x}{\sqrt{1-t}},\sqrt{\frac{1}{t}-1}\right)-4 (1-\rho )
   T\left(\frac{x}{\sqrt{t}},\sqrt{\frac{t}{1-t}}\right)+(1-\rho)
   \text{erf}\left(\frac{x}{\sqrt{2-2t}}\right)
   \right.
   \\
   \left.
   -\text{erf}\left(\frac{x}{\sqrt{2} \sqrt{t}}\right)
   \left((3
   \rho -2) \text{erf}\left(\frac{x}{\sqrt{2-2 t}}\right)+\rho -1\right)
   +3 \rho-2 -\frac{2 (\rho -1) \left(t+\sqrt{1-t}+2 \sqrt{t(1-t)}-1\right)
   e^{-\frac{x^2}{2 t}}}{\pi  \sqrt{t(1-t)}}
   \right.
   \\
   \left.
   -\frac{2 (1-\rho)
   \left(\sqrt{t}-1\right) e^{-\frac{x^2}{2 (1-t)}}}{\pi \sqrt{1-t}}\right)
   \:,
\end{multline}
where $T$ is Owen's $T$ function, defined by~\cite{Owen:1980}
\begin{equation}
  \label{eq:defOwenT}
  T(h,a) = \frac{1}{2\pi}
  \int_0^a \frac{\e^{-\frac{h^2}{2}(1+x^2)}}{1+x^2} \dd x
  \:.
\end{equation}

\subsubsection{Expressions at order \texorpdfstring{$2$}{2} in \texorpdfstring{$s$}{s}}

At second order in $s$, we do not have an explicit solution of the MFT equations~\eqref{eq:sm_MFT_mapped_back} at all times $t$. We will focus on  the final profile $\hat{q}_1^{(2)}(x,t=1)$. To determine this final profile, we actually do not need to compute $\hat{p}_1^{(2)}$.  Indeed, by defining
\begin{equation}
    \hat{q}_1^{(2)}(x,t) = -(1-\rho) \hat{p}_1^{(2)}(x,t) + Q(x,t)
    \:,
\end{equation}
the contributions of $\hat{p}_1^{(2)}$, both from the initial condition~\eqref{eq:sm_bound_pt}, and the source terms in the bulk equation~\eqref{eq:sm_MFT_mapped_back} cancel. And since at final time, $\hat{p}_1^{(2)}(x,t=1) = 0$ from the final condition~\eqref{eq:sm_bound_pt}, we do not need to determine $\hat{p}_1^{(2)}$ to compute $\hat{q}_1^{(2)}(x,t=1)$. We obtain the conditions at the origin for $Q$,
\begin{eqnarray}
    \label{eq:sm_disc_dQ}
    \partial_x Q(0^+,t) - \partial_x Q(0^-,t)
    = \frac{4(1-\rho)^3}{\rho^2} \sqrt{\frac{2}{\pi t}}
    \:,
\end{eqnarray}
\begin{multline}
    \label{eq:sm_disc_Q}
    Q(0^+,t) - Q(0^-,t)
    =
    \frac{1}{\rho }-1+\frac{4 (\rho -1)^3 t}{\pi  \rho ^2 \sqrt{(1-t)t}}
    -\frac{2 (2-\rho) (1-\rho)^2}{\pi  \rho ^2 \sqrt{1-t}}
    +\frac{4 (1-\rho)^3 \arctan\left(\sqrt{\frac{t}{1-t}}\right)}{\pi \rho^2}
    \\
    +\frac{2 (1-\rho)^2 \left(\sqrt{1-t}+2 \sqrt{(1-t) t}-1\right)}{\pi  \rho  \sqrt{(1-t) t}}
    \:.
\end{multline}
We also get a bulk equation of the form
\begin{equation}
    \partial_t Q - \frac{1}{2} \partial_{x,x} Q
    = \text{explicit source terms,}
\end{equation}
where the source terms involve the known functions $\hat{q}_0^{(n)}$, $\hat{p}_1^{(0)}$, $\hat{p}_1^{(1)}$, $\hat{q}_1^{(0)}$ and $\hat{q}_1^{(1)}$.
Since the equations are linear in $Q$, we can add the contributions of the three different equations. We first look for a solution $Q_d$ of the homogeneous equation
\begin{equation}
    \partial_t Q_d - \frac{1}{2} \partial_{x,x} Q_d
    = 0
    \:,
\end{equation}
with the conditions~(\ref{eq:sm_disc_dQ},\ref{eq:sm_disc_Q}). The solution can be written as an integral involving the heat kernel. After several manipulations, the solution at final time can be simplified into
\begin{multline}
    \label{eq:sm_Qd_for_q12}
    Q_d(x,1) = 
    \frac{\left(\pi  \left(4 \rho ^2-9 \rho +4\right)+4 (\rho -1) \rho \right)
   (\rho -1) \text{erfc}\left(\frac{x}{\sqrt{2}}\right)}{2 \pi  \rho^2}
   +\frac{(1-\rho)^3 }{\rho ^2}G(x)
   +\frac{(\rho -2) (\rho -1)^2}{\sqrt{2} \pi ^{3/2} \rho ^2} H(x)
   \\
  +\frac{(\rho -1)^2 e^{-\frac{x^2}{2}}}{\pi  \rho }
  -\frac{(\rho -1)^2}{2 \sqrt{2} \pi ^{3/2} \rho } e^{-\frac{x^2}{4}} x \mathrm{K}_0\left(\frac{x^2}{4}\right)
  \:,
\end{multline}
where  we have defined
\begin{equation}
    G(x) = \frac{1}{\pi} \sqrt{\frac{2}{\pi}}
  \int_x^\infty \e^{-z^2/4} \mathrm{K}_0 \left( \frac{z^2}{4} \right) \dd z
  \:,
  \quad
  H(x) = x \int_x^\infty \frac{\e^{-y^2/4}}{y} 
  \mathrm{K}_1 \left( \frac{y^2}{4} \right) \dd y
  - 2 \frac{\e^{-x^2/2}}{x}
  \:,
\end{equation}
with $\mathrm{K}_0$ and $\mathrm{K}_1$   modified Bessel functions.
There remains now to find the solution of the bulk equation for $Q$. We can write explicitly the solution as a space-time convolution of the source terms with the heat kernel. This allows to get a very precise numerical computation of $Q(x,t=1)$. We expect this function to be a linear combination of the functions
\begin{equation}
    \erfc \left( \frac{x}{\sqrt{2}} \right)
    \:, \:
    \erfc \left( \frac{x}{2} \right)^2
    \:, \:
    \e^{-x^2/2}
    \:, \:
    x \: \e^{-x^2/2}
    \:, \:
    x \: \e^{-\frac{x^2}{4}} \mathrm{K}_0\left(\frac{x^2}{4}\right)
    \:, \:
    G(x)
    \:, \:
    H(x)
    \:,
\end{equation}
which all appear either in the problem without bias~\cite{Grabsch:2022} or in the discontinuity computed above. We isolate the dependence of the different source terms in $\rho$ by collecting them depending on their powers of $\rho$, compute them numerically with high precision ($\sim 10^{-12}$) and fit the list of points with the functions given above. We obtain in this procedure coefficients which are either very close to integers, such as $1.0000000000000002$, or close to "reasonable" values. For instance, a prefactor of $x \: \e^{-\frac{x^2}{4}} \mathrm{K}_0\left(\frac{x^2}{4}\right)$ we obtain by this procedure is $-0.507949087468014$. Since in~\eqref{eq:sm_Qd_for_q12} this function appears with a prefactor which involves $(\pi\sqrt{2\pi})^{-1}$, we compute
\begin{equation}
    -0.507949087468014 \times (\pi\sqrt{2\pi})
    = -3.99999999995343
    \:,
\end{equation}
which we consider is $-4$ up to numerical error. By doing so with all the terms that appear, and adding the contribution of the discontinuity~\eqref{eq:sm_Qd_for_q12}, we get
\begin{multline}
  \hat{q}_1^{(2)}(x>0,1) =
  \frac{(1-2 \rho) (1-\rho)^2}{4\rho ^2} \erfc\left(\frac{x}{\sqrt{2}}\right)
  +\frac{(3-\rho)  (1-\rho)^2 }{\pi  \rho ^2} \erfc\left(\frac{x}{\sqrt{2}}\right)
  -\frac{(1-\rho)^2 }{2\rho}\text{erfc}\left(\frac{x}{2}\right)^2
  \\
  -\frac{(1-\rho)^2}{2\rho ^2} G(x)
  +\frac{5 \sqrt{2} (1-\rho)^3  }{2\pi ^{3/2} \rho ^2}x \: \e^{-\frac{x^2}{2}}
  -\frac{(3-5 \rho ) (1-\rho)^2  }{\pi  \rho ^2} \e^{-\frac{x^2}{2}}
  -\frac{\sqrt{2} (1-\rho )^2 }{2\pi ^{3/2} \rho ^2} x \: \e^{-\frac{x^2}{4}} \: \mathrm{K}_0\left(\frac{x^2}{4}\right)
   \:,
\end{multline}
and we deduce the values for $x<0$ by the replacement $x \to -x$, $s \to -s$ and $\lambda \to -\lambda$, which gives $\hat{q}_1^{(2)}(-x,1) = - \hat{q}_1^{(2)}(x,1)$.

From these results, the expression of $\Phi_1$ is deduced via relation~\eqref{eq:sm_Phi_from_qh}.

\subsection{Expressions at order \texorpdfstring{$2$}{2} in \texorpdfstring{$\lambda$}{lambda}}

We again look for a perturbative solution in $s$,
\begin{equation}
  \hat{q}_2 = \hat{q}_2^{(0)} + s \hat{q}_2^{(1)}  + \cdots
  \:,
  \quad
  \hat{p}_2 = \hat{p}_2^{(0)} + s \hat{p}_2^{(1)} + \cdots
  \:.
\end{equation}
Proceeding as for the order $1$, we find the solution
\begin{equation}
    \hat{p}_2^{(0)}(x,t) =
    \frac{(1-\rho )}{\rho}
    \left(\frac{\left(\sqrt{1-t}+\sqrt{t}+1\right)
   \e^{\frac{-x^2}{2 (1-t)}}}{2\pi\left(\sqrt{t}+1\right)}
   -T\left(\frac{x}{\sqrt{1-t}},\sqrt{\frac{1}{t}-1}
   \right)\right)
    +\frac{1}{8}
   \left(1-\text{erf}\left(\frac{x}{\sqrt{2-2 t}}\right)^2\right)
   \:,
\end{equation}
\begin{multline}
     \hat{q}_2^{(0)}(x,t) =
     \frac{(1-2 \rho ) (1-\rho )}{4 \rho }
     \left(1-\text{erf}\left(\frac{x}{\sqrt{2-2
   t}}\right) \text{erf}\left(\frac{x}{\sqrt{2} \sqrt{t}}\right)\right)
   \\
   -\frac{(1-\rho )^2 \e^{-\frac{x^2}{2 t}}}{2 \pi  \rho 
   \left(t+\sqrt{t}\right)} \left(\sqrt{t}
   \left(\left(\sqrt{1-t}+\sqrt{t}+1\right) \e^{-\frac{(2 t-1) x^2}{2 (1-t)
   t}}+\sqrt{t}+2\right)-\sqrt{1-t}-\sqrt{t(1-t)}+1\right)
   \:.
\end{multline}

At order $1$ in $\lambda$, we encounter the same problem as  for $\hat{q}_1^{(2)}$: we cannot get explicit solutions of the MFT equations. Applying the same fitting technique, with the same functions, we finally obtain
\begin{multline}
  \hat{q}_2^{(1)}(x,1) =
  \frac{(1-\rho)(1-2 \rho(1-\rho))}{4 \rho^2} \erfc \left( \frac{x}{\sqrt{2}} \right)
  + \frac{(1-\rho)^2(4-3\rho)}{2\pi \rho^2} \erfc \left( \frac{x}{\sqrt{2}} \right)
  - \frac{(1-\rho)^2}{2\rho}  \erfc \left( \frac{x}{2} \right)^2
  \\
  - \frac{(1-\rho)^2}{4 \rho^2} G(x) + \frac{2 \sqrt{2}(1-\rho)^3}{\pi^{3/2} \rho^2} x \: \e^{-\frac{x^2}{2}}
  - \frac{2(1-\rho)^2(1-2\rho)}{\pi \rho^2} \e^{-\frac{x^2}{2}}
  - \frac{(1-\rho)^2}{2\sqrt{2} \pi^{3/2} \rho^2}  x \: \e^{-\frac{x^2}{4}} \: \mathrm{K}_0 \left( \frac{x^2}{4} \right)
  \:.
\end{multline}

From these results, the expression of $\Phi_2$ is deduced via relation~\eqref{eq:sm_Phi_from_qh}.

\subsection{Obtaining the cumulants}

Using the boundary relation~\eqref{eq:sm_BoundPhi0}, we can write the cumulant generating function as
\begin{equation}
    \hat{\psi} = \mp (\e^{\pm \lambda}-1) \frac{\Phi'(0^\pm)}{2\Phi(0^\pm)}
    = \sum_{n \geq 1} \hat\kappa_n \frac{\lambda^n}{n!}
    \:.
\end{equation}
By doing so, we first check that, using the expressions of $\Phi$ obtained above, these two equations (either with $\Phi'(0^+)$ or with $\Phi'(0^-)$) give the same expression for $\hat\psi$. Consequently, this gives the cumulants
\begin{equation}
  \hat\kappa_2 = \frac{1-\rho}{\rho \sqrt{\pi}}
  +\frac{(1-\rho)^2 \left(7-5 \rho -\pi  \left(\left(\sqrt{2}-3\right) \rho
   +2\right)\right)}{\pi ^{3/2} \rho ^3} s^2 + O(s^3)
  \:,
\end{equation}
\begin{equation}
  \hat\kappa_3 = s \frac{ (1-\rho ) \left(
      12 (1-\rho )^2
      -\pi  \left(\left(8-3 \sqrt{2}\right) \rho ^2-3 \left(4-\sqrt{2}\right) \rho +3\right)
    \right)}{\pi ^{3/2} \rho ^3}
  + O(s^2)
\end{equation}

\section{Numerical simulations}

The simulations of the SEP are performed using the Kinetic Monte Carlo method \cite{Schulze2008}. We consider a periodic ring of size $N$, where initially on each site (except site 0) we place a particle with probability $\rho \in [0,1]$, the average density. At site 0, we systematically place a particle, which will be the tracer particle. Then, we draw an integer $E$ from a Poisson distribution of parameter $N\times t$ where $t$ is the time at which we want to sample our system. The integer $E$ is the number of jumps that will be attempted between initial time and $t$. Therefore we perform $E$ times the following procedure : we pick a particle at random ; if it is not the tracer, we pick a direction (left or right) with probability 0.5, if the corresponding site is empty, the jump is performed, otherwise, nothing happens ; if it is the tracer, we pick the direction right with probability $\frac{1+s}{2}$ and left with probability $\frac{1-s}{2}$, if the corresponding site is empty, the jump is performed, otherwise, nothing happens. 

During the procedure, we keep track of the position of the tracer $X_t$, and we average $X_t^k$ over about $10^7$ simulations in order to get an estimate of the $k$-th moment of the position of the tracer particle. To get the generalized density profiles in the reference frame of the tracer, we average $\eta_{X_t + r}X_t^k$. Typically we choose the number of sites $N = 2000$ or $3000$ depending on the final time so that the values sampled does not depend on $N$ anymore.


%